\documentstyle[12pt,preprint,prd,aps,epsf,floats]{revtex}
\tightenlines
\begin{document}
\input psfig
\draft
\begin{titlepage}

\title{\bf  Is Large Mixing Angle 
MSW the Solution of the Solar Neutrino Problems? }
\author{J. N. Bahcall}
\address{School of Natural Sciences, Institute for Advanced
Study\\
Princeton, NJ 08540\\}
\author{P. I. Krastev}
\address{University of Wisconsin--Madison, Department of Physics, \\
Madison WI 53706}
\author{A. Yu. Smirnov}
\address{International Center for Theoretical Physics, 34100 Trieste, Italy}
\maketitle
\begin{abstract}
Recent results on solar neutrinos 
provide hints that the LMA MSW
solution could be correct.
We  perform accurate  calculations for potential   `smoking-gun'  effects 
of the LMA solution  
in the SuperKamiokande solar neutrino experiment, including: 
(1) an almost constant reduction of the standard 
recoil electron energy spectrum
(with a weak, $< 10$\%, relative increase below $6.5$ MeV);
(2) an integrated difference in day-night rates ($2$\% to $14$\%);
(3) an approximately constant zenith-angle dependence 
of the nighttime event rate;
(4) a  new test for the 
difference in the shape of the equally-normalized day-night
energy spectra ($\sim 1$\%); 
and (5) annual variations of the signal due to the regeneration effect
($\sim 6$ times smaller than the integrated day-night effect).
We also establish a relation between the integrated day-night asymmetry and
the seasonal asymmetry due to LMA regeneration.
As a cautionary example, we simulate the effect of an absolute energy
calibration error on the shape (distortion) of the recoil energy
spectrum.

We compare LMA  predictions with the available SuperKamiokande data
and  discuss the possibilities for
distinguishing experimentally between LMA and vacuum oscillations.
If LMA is correct, 
global solutions combining data from different types of measurements made by
SuperKamiokande or by different solar
neutrino experiments could reveal in the next few years a many $\sigma$
indication of neutrino oscillations.
\end{abstract}
\end{titlepage}
\newpage

\section{Introduction}
\label{sec:introduction}
On April 1, 1996, solar neutrino research shifted from the pioneering
phase of the chlorine (Homestake) experiment~\cite{chlorine} 
and the exploratory phase of the
Kamiokande~\cite{kamiokande}, SAGE~\cite{SAGE},  and
GALLEX~\cite{GALLEX}  
experiments, to the era of precision
measurements inaugurated by the 
SuperKamiokande~\cite{superkamiokande300,neutrino98,superkamiokandeDN,superkamiokande504,superkamiokande708,suzuki99,minikata}
 experiment.   In this
paper, we concentrate on the predictions for `smoking-gun' evidence of
new neutrino physics that can be observed in electron-neutrino
scattering experiments like SuperKamiokande.

There are
four generic oscillation solutions
(LMA, SMA, and LOW MSW solutions and vacuum oscillations,
cf. Ref.~\cite{bks98} for a recent discussion)
of the solar neutrino problems that 
involve two neutrinos. These
solutions are all consistent with the predictions of the standard
solar model~\cite{BP98} and the observed average event  rates  in the
chlorine, Kamiokande, SuperKamiokande, GALLEX, and SAGE experiments.
We limit ourselves to such globally consistent 
oscillation solutions and adopt the methodology used in our recent
paper~\cite{bks98}. 

Recent experimental results from
SuperKamiokande~\cite{neutrino98,superkamiokandeDN,superkamiokande504,superkamiokande708}
provide some encouragement for considering the LMA solution of the MSW
effect~\cite{msw}. No statistically 
significant distortion of the recoil electron
energy spectrum has been discovered. No 
obvious change has been revealed in the slope of 
 the ratio  [Measurement/(Standard Spectrum)] as 
the SuperKamiokande measurements have been extended to progressively
lower energies.  Moreover, the excess of
events at the highest energies has diminished. 

A larger flux has been
observed at night than during the day, although the difference 
observed by
SuperKamiokande~\cite{neutrino98,superkamiokandeDN,superkamiokande504,superkamiokande708}
 (which
is about the size of what is expected from the LMA solution) is only 
$\sim 1.6\sigma$ after $708$ days of data collection~\cite{inoue99}.
In addition,  no large enhancement of the rate has been found when
the neutrino trajectories cross the earth's core. Moreover, the excess
that is found appears already when the sun is just below the horizon.
These results
are, as we shall see in detail in this paper, expected on the basis of
the LMA MSW solution but are not expected on the basis of the SMA solution.

There are also some non-solar indications that large mixing angles may
be plausible. First, and most important, SuperKamiokande and other
experiments on atmospheric neutrinos have provided
strong evidence that large lepton mixing is occurring at least in one
channel involving muon 
neutrinos~\cite{fukuda98}. Moreover, the mass required for the 
LMA solution ($m
\sim \sqrt(\Delta m^2) \sim 10^{-2} {\rm eV} $) is only one order of
magnitude smaller than the mass scale suggested by atmospheric
neutrino oscillations. All of the other solar neutrino solutions imply
a smaller or 
much smaller neutrino mass scale for solar neutrinos. It is however
well known that a weak mass hierarchy allows a rather natural
explanation of large mixing.

The oscillation parameters of the LMA solar neutrino solution
can give an observable effect for atmospheric neutrino experiments. In
particular, the LMA solution could explain~\cite{smirnov99} 
the excess of $e-$like
events in the atmospheric data.

In this paper, we provide accurate calculations for smoking-gun
predictions of the LMA solution, including a simple new test
(Sec.~\ref{subsec:spectrumsolar}). In particular, 
we calculate and compare with SuperKamiokande observations the
following quantities: (1) the distortion of the electron recoil energy
spectrum; (2) the zenith angle dependence of the observed counting
rate; (3) the day-night spectrum test; and 
 (4) the seasonal dependence (beyond the effect of the earth's
eccentric orbit) of the counting rate.  All of
these quantities are zero, independent of solar physics, if the standard
model for electroweak interactions is correct. Therefore, measurements of the
spectral energy distortion and the zenith angle and seasonal dependences
constitute sensitive  smoking-gun tests of new physics.
As a cautionary example, we also simulate the effect of an energy
calibration error on the apparent distortion of the electron recoil
energy spectrum.

\begin{table}[!t]
\baselineskip=16pt
\caption[]{Standard Solar Model and LMA Predictions for Solar Neutrino 
Rates.
The standard solar model predictions are taken from BP98, Ref.~\cite{BP98}.
The LMA predictions refer to the best-fit LMA solution when only the
total rates are considered: $\Delta m^2 = 2\times 10^{-5}
{\rm eV^2}$ and $\sin^2 2\theta = 0.8$ (cf. Ref.~\cite{bks98}).
[The values given in parentheses were computed using, respectively,
the oscillation parameters:
$\Delta m^2 = 8 \times 10^{-5}~{\rm eV^2},
\sin^2 2\theta = 0.63$ and $\Delta m^2 = 8 \times 10^{-5}~{\rm eV^2},
\sin^2\theta = 0.91$.]
The rates calculated using the BP98 fluxes and the LMA oscillation
solution agree with the measured rates for
the Homestake 
chlorine  experiment ($2.56 \pm 0.23$ SNU, see Ref.~\cite{chlorine}) and the 
SAGE ($67 \pm 8$ SNU, see
Ref.~\cite{SAGE}) and GALLEX
($78 \pm 6$ SNU, see Ref.~\cite{GALLEX})  gallium experiments.
\protect\label{tab:LMArates}}
\begin{tabular}{llcccc}
Source&\multicolumn{1}{c}{Flux$^a$}&Cl&Ga&Cl&Ga\\
&&(SNU)&(SNU)&(SNU)&(SNU)\\
&\multicolumn{1}{c}{Standard}&{Standard}&{Standard}&{LMA}&{LMA}\\
\noalign{\smallskip}
\hline
\noalign{\smallskip}
pp&$5.94 $&0.0&69.6&0.0 (0.0,0.0)&39.8 (47.1,37.5)\\
pep&$1.39 \times 10^{-2}$&0.2&2.8&0.1 (0.1,0.1)&1.2 (1.8,1.4)\\
hep&$2.10 \times 10^{-7}$&0.0&0.0&0.0 (0.0,0.0)&0.0 (0.0,0.0)\\
${\rm ^7Be}$&$4.80 \times 10^{-1}$&1.15&34.4&0.46 (0.65,0.60)&15.7 (21.5,17.4)\\
${\rm ^8B}$&$5.15 \times 10^{-4}$&5.9&12.4&1.8 (1.6,2.3)&3.7 (3.4,4.8)\\
${\rm ^{13}N}$&$6.05 \times
10^{-2}$&0.1&3.7&0.05 (0.06,0.05)&1.8 (2.4,1.9)\\
${\rm ^{15}O}$&$5.32 \times
10^{-2}$&0.4&6.0&0.16 (0.25,0.20)&2.5 (3.8,3.1)\\
\noalign{\medskip}
&&\hrulefill&\hrulefill&\hrulefill&\hrulefill\\
Total&&$7.7^{+1.2}_{-1.0}$&$129^{+8}_{-6}$&2.6 (2.65,3.2)&64.7 (80,66)\\
\noalign{\smallskip}
\end{tabular}
\hbox to\hsize{$^{\rm a}$ All fluxes are in units of 
$\left(10^{10}\ {\rm cm^{-2}s^{-1}}\right)$\hfill}
\end{table}

\begin{figure}
\centerline{\epsfxsize=5.5in\epsffile{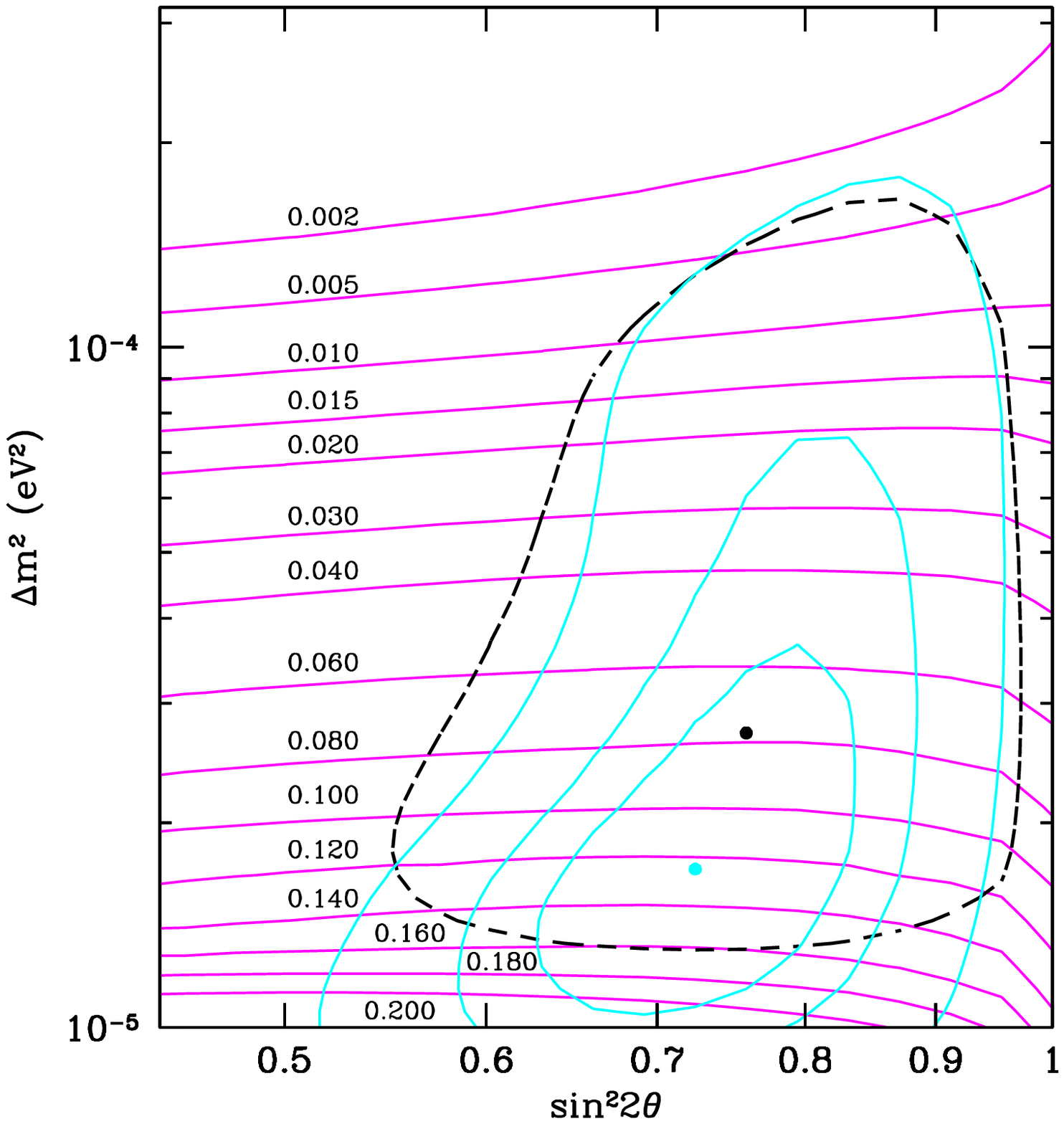}}
\caption[]{ 
The allowed region of the LMA MSW parameter space.
When only the rates in the chlorine, SuperKamiokande, SAGE, and GALLEX
experiments are considered, 
the vertical continuous lines are part of the allowed contours at
$90$\%,  $95$\%, and $99$\% C.L. 
The dashed contour corresponds to the allowed region at $99$\%
C.L. when both the total rates and the Night-Day asymmetry are
included. 
The best fit parameters, indicated by a dot in the figure,
are for the rates-only fit: $\sin^2(2\theta) =
0.72$ and $\Delta m^2 = 1.7\times 10^{-5} {\rm eV^2}$. 
The best-fit
parameters for the combined fit are
$\sin^2(2\theta) =
0.76$ and $\Delta m^2 = 2.7\times 10^{-5} {\rm eV^2}$. 
The approximately horizontal lines show the contours for different
Night-Day asymmetries (numerical values indicated); see
Ref.~\cite{guth99} for an illuminating discussion of the Night-Day
asymmetry. 
The data are from Refs.~\cite{chlorine,SAGE,GALLEX,superkamiokande708}.
}
\label{fig:zoom}
\end{figure}

Figure~\ref{fig:zoom} shows the allowed region of LMA solution
 in the plane of $\sin^2 2\theta$
and $\Delta m^2$.
We have indicated 
the region allowed by the measured total rates in solar neutrino experiments
 by the continuous contours in Fig.~\ref{fig:zoom}; the contours  
are shown at $90$\%, $95$\%, and $99$\% C.L. 

The constraints provided by
the  total rates in the Homestake, GALLEX, SAGE, and
SuperKamiokande experiments 
are, at present, the most robust limitations on the allowed
parameter space. 
The only slight 
change in the available data on total rates
since the publication of Ref.~\cite{bks98} is a 
small reduction after 708 days in the total error for the 
SuperKamiokande experiment~\cite{superkamiokande708}; the
allowed region based upon the total rates in solar neutrino
experiments is essentially unchanged from our earlier study
(cf. the similar results 
in Fig.~2 of Ref.~\cite{bks98} for the case
that only the total rates are considered).

The dashed line in Fig.~\ref{fig:zoom} represents the  
$99$\% allowed contour including both the measured SuperKamiokande day-night
effect and the total rates.
The SuperKamiokande collaboration has performed preliminary global
analyses of the available $708$ days of
data~\cite{superkamiokande708,suzuki99,inoue99} .

Table~\ref{tab:LMArates} contrasts the predictions of the standard
model and the LMA for the chlorine and the gallium experiments. 
The LMA results that are not in parentheses are  for 
$\Delta m^2 = 2\times 10^{-5}
{\rm eV^2}$ and $\sin^2 2\theta = 0.8$.  
These neutrino parameters
are close to the best-fit LMA values when only the total rates of the 
chlorine, Kamiokande, SuperKamiokande, GALLEX, and SAGE 
experiments are considered (see Ref.~\cite{bks98}).
In parentheses, we show the predictions for two rather extreme LMA
solutions.  We have evaluated the predictions for the chlorine and
gallium experiments for a representative set of oscillation parameters
and all of the values lie within the boundaries defined by the three
solutions shown in Table~\ref{tab:LMArates}. Therefore, the
predictions given in the table show the expected range of capture
rates consistent with the LMA solution and the existing experiments.
The range of production rates predicted by the set of LMA oscillation
solutions considered in this paper is: $^{37}{\rm Cl} = 2.9 \pm 0.3$
SNU and $^{71}{\rm Ga} = 71 \pm 9$ SNU.

The prediction of the LMA solution shown in Table~\ref{tab:LMArates}
for the SuperKamiokande rate due to  the $^8$B
neutrino
flux~\cite{superkamiokande300,neutrino98,superkamiokande504,superkamiokande708}
is :\
${\rm
Rate\ ({\rm measured}, ^8B)
= 0.474\  Rate\ ({\rm BP98}, ^8B)}$.

This paper is organized as follows. We discuss the distortion of the
recoil electron energy spectrum in Sec.~\ref{sec:spectrum}, the
day-night differences of the event rate in
Sec.~\ref{sec:zenithdependence}, and the seasonal dependences in
Sec.~\ref{sec:seasonal}.  
We describe in Sec.~\ref{sec:vacuumvsLMAseasonal} 
possibilities for distinguishing between LMA and vacuum
oscillations. 
We summarize and discuss our main results in Sec.~\ref{sec:discussion}.

\section{Spectrum distortion}
\label{sec:spectrum}

Figures~\ref{fig:energyerror} and \ref{fig:specconsttheta}  
illustrate the main results of
this section. 
The reader who wants to get quickly the essence of the physical
situation can skip directly to the discussion of
Fig.~\ref{fig:energyerror} in Sec.~\ref{subsec:energyerror} and 
Fig.~\ref{fig:specconsttheta} in
Sec.~\ref{subsec:predictedvsspectra}.

The observation of a distortion 
in the recoil electron energy spectrum produced by solar neutrinos
scattering off electrons 
would be a definitive signature of physics beyond the 
Standard Electroweak Model.
Neutrinos which correspond to recoil electrons 
with energies between $5$ MeV and $13$ MeV,
which are most easily observed by 
 the SuperKamiokande solar neutrino experiment
~\cite{superkamiokande300,neutrino98,superkamiokande504,superkamiokande708}, 
are predicted by 
the Standard Solar Model to be essentially all (more than 99\%)  
from $^8$B beta decay~\cite{BP98}. 
The shape of the $^8$B neutrino energy 
spectrum can be determined accurately 
from laboratory measurements~\cite{balisi} and the influence of 
astrophysical factors 
is less than  $1$ part in $10^5$ \cite{bahcall91}. 
For energies relevant to solar neutrino studies, 
the cross-sections for neutrino scattering have been calculated 
accurately 
in the standard electroweak model, including radiative
corrections~\cite{sirlin}.  
Therefore, if nothing happens to solar neutrinos 
on the way from the region of production in the solar interior 
to a  detector on earth, 
the recoil electron energy spectrum from $^8$B neutrinos can be
calculated precisely.

Any measured deviation of the recoil electron energy spectrum 
from  the standard  shape  would indicate new physics. 
The opposite, however, is not true: 
the absence of a measured  distortion does not mean the absence of 
new physics.

We first indicate in Sec.~\ref{subsec:convolution} how
the neutrino energy spectrum is convolved with the electron recoil
energy spectrum and with the measurement characteristics of the
detector. We then summarize  briefly in Sec.~\ref{subsec:summaryofdata} the
characteristics of the measured SuperKamiokande energy spectrum.
In Sec.~\ref{subsec:energyerror}, we answer the question: How does an
error in the energy calibration affect the shape of the recoil energy
spectrum? 
We discuss in Sec.~\ref{subsec:predictedvsspectra} the comparison
of the predicted and the measured electron energy spectra, which are
presented in Fig.~\ref{fig:specconsttheta}.

\subsection{The convolution}
\label{subsec:convolution}

We concentrate in this paper on  analyzing the ratio, $R(E_i)$, 
of the number of observed events, $N^{\rm Obs}_i$ 
in a given  energy bin, $E_i$, to 
the number, $N^{\rm SSM}_i$,
 expected from the SSM~\cite{BP98}, where  

\begin{equation}
R(E_i) = \frac{N^{\rm Obs}_i}{N^{\rm SSM}_i} ~.
\label{eq:Rdefinition}
\end{equation}
For specific applications, 
we use the observed event numbers  
in the form provided by 
SuperKamiokande~\cite{superkamiokande504,superkamiokande708}; there
are $17$  energy bins of $0.5$ MeV width 
from $5.5$ MeV to $14$ MeV and an $18$th
bin that includes all events from $14$ MeV to $20$ MeV.
When more data are available, it will be important to 
divide the events above $14$ MeV into several different bins~\cite{bk98}.

The number of events in a given energy bin, $i$, can be expressed as 
$$
N_i = \int_{E_i}^{E_i + 0.5 MeV} dE_e 
\int^{\infty}_0 dE_e' f(E_e, E_e')
\int^{\infty}_{E_e'} dE_{\nu} F(E_{\nu})\times  
$$
\begin{equation} 
\left[ P(E_{\nu}) \frac{d\sigma_e (E_{\nu}, E_e') }{d E_e'} 
+ (1 - P(E_{\nu})) \frac{d\sigma_{\mu} (E_{\nu}, E_e') }{d E_e'} \right], 
\label{numberi}
\end{equation}
where 
$F(E_{\nu})$ is the flux of $^8$B  neutrinos 
per unit energy at the detector, 
$f(E_e, E_e')$ is the energy resolution function which 
we take from~\cite{superkamiokande300,neutrino98}.  
$P(E_{\nu})$ is the survival probability 
$\nu_e \rightarrow \nu_e$, 
$d\sigma_e/d E_e'$ and $d\sigma_{\mu}/d E_e'$ are the differential 
cross-sections of the $\nu_e e - $ and $\nu_{\mu} e - $ scattering. 
There are practically  no $^8$B neutrinos with energies 
greater than  $15.5$ MeV, so that for practical purposes the upper
limit of the integral over $E_\nu$ in Eq.~(\ref{numberi}) can be taken
to be $16$ MeV. 
Including the effects of the finite energy resolution,
the upper limit over  the measured electron recoil
energy, $E_e'$, can be taken to be $20$ MeV.

To use Eq.~(\ref{numberi}) to calculate the combined prediction for
the standard  electroweak model and the standard solar model, set 
$P \equiv 1 $. The second 
 term in the brackets of Eq.~(\ref{numberi}) then disappears. This
term is also absent if the electron neutrinos are converted to sterile
neutrinos.

How does the convolution shown in Eq.~(\ref{numberi}) affect
energy-dependent distortions (or conversions) of the incoming electron
neutrinos?  The convolution spreads out over a relatively wide energy
range any energy-dependent features. In particular, 
the scattering process produces electrons with energies from zero to
almost equal to the incoming neutrino energy (for neutrino energies
much greater than the electron mass). 
If  $\nu_e$ are converted to 
$\nu_{\mu}$ or/and $\nu_{\tau}$, then there will be  neutral current
scattering [the second term in boxed brackets of Eq.~(\ref{numberi})]
which, while less probable, will also reduce the apparent effect of
the conversion.
Since the solar $^8$B neutrino
spectrum decreases rapidly beyond about $10$ MeV, 
a  distortion of the recoil electron spectrum 
at some energy $E_{e}$ 
($E_{e} > 10$ MeV) is determined by the distortion of the neutrino
spectrum at  somewhat lower energy $E_{\nu} <  E_{e}$.
The energy resolution function has a crucial
affect in smearing out distortion. The $2\sigma$ width for the energy
resolution is about $3$ MeV at $E_e = 10$ MeV in
SuperKamiokande~\cite{superkamiokande300,neutrino98,superkamiokande504}. 
Smaller features will be smeared out.

\subsection{Summary of the data}
\label{subsec:summaryofdata}

After $708$ days of data taking with  SuperKamiokande, 
no unequivocal distortion of the electron recoil energy spectrum 
has been found.
The data 
show essentially the spectrum shape expected for no distortion 
for  $E < 13$ MeV with  some excess events at 
$E > 13$ MeV. 
The excess electrons observed at higher energies may reflect: 
(1) a statistical fluctuation; (2) the contribution of $hep$ 
neutrinos~\cite{bk98,escribano,fiorentini}; 
(3) a larger-than-expected error in the absolute energy normalization;
or  (4) new physics.
The systematic effects due to $hep$ neutrinos or to an error in the
energy normalization can hide a distortion due to neutrino conversion
or cause an artificial distortion.
For example, a  suppression due to oscillations
of the high energy part of the electron
spectrum could  be compensated for (hidden by)  the effect of 
{\it hep} neutrinos. 

All of the four explanations for the high-energy excess have been
described previously in the literature. However, the distortion due to
an error in the absolute energy scale of the electrons 
has only been mentioned as a
possibility~\cite{bl96}. Therefore, we discuss 
in Sec.~\ref{subsec:energyerror}
the effect on the energy spectrum of an error in the absolute energy scale.

Another possible systematic error,
a non-Gaussian tail to the energy resolution, could also produce an
apparent distortion at the highest energies. However, this effect has
been estimated by the SuperKamiokande collaboration and found to be
small~\cite{liu,superkamiokande708}. 

To resolve the ambiguity at high recoil electron 
energies,  one can  proceed in at least two  
different ways. 
(1) One can exclude the spectral data above $13$ MeV,
since the distortion due to both $hep$ neutrinos 
and the calibration of the energy scale normalization occur mainly at high 
energies.
A comparison of the results of the analysis using all the data with
the inferences reached using only data below $13$ MeV provides an estimate
of the possible influence of systematic uncertainties. 
However, this procedure does throw away some important data.
(2) One can treat the solar $hep$ neutrino flux as a free parameter 
in analyzing the energy spectrum~\cite{bk98,escribano}. 
In general, this is the
preferred method of analysis and the one which we follow.

\subsection{How does an energy calibration 
error affect the recoil energy spectrum?}
\label{subsec:energyerror}

An error in the 
absolute normalization of the energy of the recoil electrons,  
$\delta = E_{\rm true} -E_{\rm measured}$,  would lead to an
apparent energy dependence in  $R$ that is particularly important for
the LMA solution, since the theoretical prediction is that $R$ is
essentially constant in the higher energy region accessible to
SuperKamiokande. The ratio
$R(E)$ can be written as 
\begin{equation} 
R^{\rm Obs}(E) ~= ~
\frac{N^{\rm Obs}(E + \delta)}{N^{\rm SM}(E)}
~\approx~
R_0^{\rm true}(E) \left( 1  + \frac{1}{N^{\rm SM}(E)} \frac{dN^{\rm SM}(E)}{dE} 
\delta \right)~,  
\label{eq:absolute}
\end{equation}
where $N^{\rm SM}(E)$ is the standard model spectrum.
Since 
$N^{\rm SM}$, $dN^{\rm SM/}dE$ and, in general, 
$\delta $ all depend on energy, 
the last term in  Eq.~(\ref{eq:absolute}) gives rise to an apparent
distortion of the observable spectrum. 
Moreover,  since $N^{\rm SM}(E)$ is a decreasing function of energy, 
$\delta \times (dN^{\rm SM}/dE)/N^{\rm SM}$ 
increases with energy for negative 
$\delta $ . Thus, for negative $\delta $,  
$R(E)$ is  enhanced at high energies.
The SuperKamiokande collaboration 
has estimated the error in the absolute energy
calibration as being $0.8\%$  at 10 MeV~\cite{superkamiokande504}, i.e.,  
$\delta  = 80~~ {\rm keV}$, $1\sigma$.

We have simulated the effect of an error in the absolute energy
calibration by convolving a standard model recoil energy spectrum with the
energy resolution function of SuperKamiokande and introducing a constant
offset error $\delta$. We calculated the ratio $R(E)$ directly from
the equality in Eq.~(\ref{eq:absolute}), without making the
approximation involved in the Taylor-series expansion.

Figure~\ref{fig:energyerror} compares the computed $R(E)$ for three
different values of $\delta$ with the
observed SuperKamiokande data~\cite{superkamiokande708} for $708$
days. 
For visual convenience, each  curve is  normalized somewhat
differently, but all are normalized 
within the range allowed by the SuperKamiokande
measurement of the total rate. We see from Fig.~\ref{fig:energyerror}
that a  $2\sigma$ ($160$ keV) energy calibration error can produce a
significant pseudo-slope.
More exotic effects can be produced if the one assumes that the energy
error, $\delta$, is itself a function of energy.

We conclude that future analyses of the distortion of the 
energy spectrum should include $\delta$ as one of the
parameters that is allowed to vary in determining from the
measurements the best-fit and the
uncertainty in the energy distortion.

\subsection{Predicted versus measured spectra}
\label{subsec:predictedvsspectra}

Figure~\ref{fig:specconsttheta} compares the predicted versus the measured
ratio $R(E)$ [see Eq.~(\ref{eq:Rdefinition})] of the 
electron recoil energy spectra for a representative 
range of $\Delta m^2$
that spans  the domain permitted by the global
LMA solutions.  All of the solutions shown 
have $\sin^2 2\theta = 0.8$ and are 
 consistent with the average event rates in the
chlorine, Kamiokande, SAGE, GALLEX, and SuperKamiokande experiments.
Also, all of the  histograms of neutrino solutions are
normalized so as to yield the same value of the ratio $R(E)$ at $E = 10$
MeV. 
The measured values are taken from~\cite{superkamiokande708}.

The most conspicuous feature of Fig.~\ref{fig:specconsttheta} 
is the flatness of the  ratio
$R(E)$ that is predicted by the LMA solution with 
 no enhancement of the $hep$ flux. The survival probability, $P$,
without earth regeneration is  
 $P(E) \approx \sin^2 \theta$ in the observable high energy part of the $^8$B
neutrino spectrum. Regeneration in the earth can lead to a weak
increase of the survival probability with energy.

Excellent fits  to the recoil energy spectra can be obtained by
increasing the production cross section for the 
$hep$ flux by a factor of $10$ to $40$ over its nominal (see Ref.~\cite{adelberger98})
best value. 
This result is illustrated in 
Fig.~\ref{fig:specconsttheta}.

Results similar to Fig.~\ref{fig:specconsttheta} are obtained if 
 $\sin^2 2\theta$ is allowed to vary  over
a representative range of  the allowed global LMA solutions that are
consistent with the average  measured event rates. 
We find numerically that the dependence of the shape of the predicted
spectrum upon $\sin^2 2\theta$ is very weak.

Recent work by the SuperKamiokande collaboration has
placed an experimental upper limit on the $hep$ neutrino flux at earth
of at
most $20$ times the nominal standard
value~\cite{neutrino98,superkamiokande708,suzuki99}.  The fits shown in
Fig.~\ref{fig:specconsttheta} demonstrate that 
 we require a $hep$ flux at the sun of 
$10$ to $40$ times the standard value. However, LMA neutrino oscillations
reduce the rate  at the earth so that the observed rate is about a
factor of two less than would correspond to a flux of purely  $\nu_e$ 
$hep$ neutrinos. So the fits shown for variable $hep$ flux 
in Fig.~\ref{fig:specconsttheta} and are consistent with the SuperKamiokande
upper limit on the measured $hep$ flux.
 
What are the chances that SuperKamiokande can obtain 
 smoking-gun evidence for a
distortion of the energy spectrum if the LMA solution is correct?  The lack
of knowledge of the $hep$ production cross section prevents
fundamental conclusions based upon the suggestive higher energy ($>
 13$ MeV) events. The data reported by SuperKamiokande 
(at the time this paper is being
 written)  above $14$ MeV 
are in a single bin ($14$ MeV to $20$
 MeV). As emphasized in Ref.~\cite{bk98}, it is possible that
 measurements in smaller energy bins above $14$ MeV could 
 reduce the uncertainty in the $hep$ flux and test the
 consistency of the LMA plus enhanced $hep$ predictions (for
 preliminary results see Ref.~\cite{superkamiokande708}).

For larger allowed values of  $\Delta m^2$, the distortion curve turns up at
low energies 
(cf. Fig.~\ref{fig:specconsttheta}). This upturn is 
due to the fact that for larger
$\Delta m^2$ the low energy part of the $^8$B neutrino spectrum 
($E \sim  5 - 6 $ MeV) is on the adiabatic edge of
the suppression pit. 
For the smallest allowed  
$\Delta m^2$, LMA predicts a  weak positive 
slope for $R(E)$ that is caused by the  Earth regeneration effect.

The predicted distortion at low energies, between, e.g., $5$ MeV and 
 $6.5$ MeV, is  small (typically $\sim$ a few  
percent up to as large as $10$ percent, cf. Fig.~\ref{fig:specconsttheta}) 
and would require approximately $10$ years
 of SuperKamiokande data in order to show up in a clear way.

\section{Earth regeneration effects}
\label{sec:zenithdependence}

If MSW conversions occur, the sun is predicted to be brighter at
night in neutrinos than it is during the
day~\cite{earthreg} at those energies at which the survival
probability of $\nu_e$ is less than one-half.
This phenomenon, if
observed, would  be a dramatic smoking-gun indication of new
physics independent of solar 
models\footnote{If neutrino oscillations occur, the
 predicted zenith-angle dependence changes slightly for different solar
models since the flavor content of the calculated solar flux arriving
at earth is influenced somewhat by the assumed standard model
fluxes. See Ref.~\cite{brighter}, Sec. XB.}.

When the sun is below the horizon, neutrinos must traverse part of the
earth in order to reach the detector. By interacting with terrestrial
electrons, the more difficult to detect $\nu_\mu$ or $\nu_\tau$ can be
reconverted to the more easily detectable $\nu_e$. The opposite
process can also occur. The physics of this `day-night' effect is well
understood~\cite{earthreg}. We adopt the methodology of
Ref.~\cite{brighter}.  The SuperKamiokande collaboration has placed
constraints on neutrino oscillation parameters from the measurement of
the zenith angle dependence in $504$ days of data~\cite{superkamiokande504}.

We first discuss in Sec.~\ref{subsec:integrated} the 
day-night asymmetry averaged over all
zenith angles and integrated over all energies and seasons of the
year. We then analyze in 
 Sec.~\ref{subsec:zenith} the 
dependence of the total rate upon the solar zenith angle. 
We then describe in  
 Sec.~\ref{subsec:spectrumsolar} a new test, which we
call the `day-night' spectrum test.

\subsection{The Integrated Day-Night  Effect}
\label{subsec:integrated}

The average day-night asymmetry measured by the SuperKamiokande
collaboration is~\cite{inoue99}

\begin{equation}
A_{\rm n-d} ~=~ 2\left[{\rm  {\frac{night - day}{night + day} } } \right] 
 ~=~  0.060 \pm 0.036({\rm stat}) \pm 0.008 ({\rm syst}).
\label{eq:daynightasym}
\end{equation}
Here `night' (`day') is the  nighttime (daytime) signal averaged
over energies above $6.5$ MeV and 
averaged over all zenith angles and seasons of the
year.\footnote{The definition of $A_{\rm n-d}$ given in
Eq.~(\ref{eq:daynightasym}) is twice as large as in
Ref.~\cite{brighter}; we changed our definition to the one given in
the present paper in order to conform to
the usage by the SuperKamiokande collaboration. Similarly, we do not
discuss the valuable constraints provide by moments of the day-night
effect because the SuperKamiokande collaboration has not yet provided
an analysis of their results in terms of moments.}

Table~\ref{tab:daynight} shows the calculated values for the day-night
asymmetry for a range of different LMA solutions.
For solutions within the allowed LMA domain, we find
$A_{\rm n-d} = 0.02$ to $0.14$ .

The dependence of 
the asymmetry on 
$\Delta m^2$ can be approximated with the LMA solution space by 
\begin{equation}
A_{\rm n-d} ~=~ 0.2 \left[\frac{10^{-5} {\rm eV}^2}{\Delta m^2}
\right]
\label{eq:daynightapprox}
\end{equation}
for $\sin^2 2\theta = 0.8$ and for the range of $\Delta m^2$ shown
in Table~~\ref{tab:daynight}.
This approximation depends only weakly on mixing angle. 
The $1 \sigma$ interval of $A_{\rm n-d}$ that follows 
from Eq. (\ref{eq:daynightasym}) 
leads to an allowed range of
\begin{equation}
\Delta m^2 = (2 - 8) \times 10^{-5} {\rm eV}^2, ~~~~~ 1 \sigma~. 
\label{eq:daynightmass}
\end{equation}

Figure~\ref{fig:zoom} shows the approximately horizontal 
lines of equal Night-Day asymmetry 
in a $\Delta m^2$ - $\sin^2 2\theta$ plot  together with 
region in  parameter space (dashed contour) 
that is   allowed by a combined fit of the 
 total rates
and the Night-Day asymmetry.  
The best-fit oscillation
parameters for the combined fit are

\begin{equation}
\label{bestall}
\Delta m^2  =  2.7\times 10^{-5} {\rm eV}^2 ;
~\sin^22\theta  =  0.76 .
\end{equation}
The Night-Day asymmetry is $8$\% for these
best fit parameters.

\subsection{The Zenith Angle Dependence}
\label{subsec:zenith}

Figures~\ref{fig:zenith} and \ref{fig:zenithconstm}
compare the predicted and the observed (after
$708$ days) 
dependence of the 
SuperKamiokande~\cite{superkamiokande708} event rate upon the zenith
angle of the sun, $\Theta$. The results are averaged over one year.  
In order to make the figures look most similar to figures published by
the SuperKamiokande collaboration, we have constructed the plot using
the nadir angle $\Theta_N = \pi - \Theta$.
The predicted and the observed rates
are averaged over all energies $> 6.5$ MeV.

Three conclusions can be drawn from the results shown in 
Fig.~\ref{fig:zenith} and Fig.~\ref{fig:zenithconstm}. 
(1) The experimental error bars must 
be reduced by about a
factor of two (total observation time of order eight years) before one
can make a severe test of the average zenith-angle distribution predicted by
the LMA solution. (2) Nevertheless, the available experimental 
results provide a
hint of an effect: all five of the nighttime rates are larger than the
average rate during daytime. (3) The zenith-angle dependence predicted by the
LMA solution is relatively flat, i.e.,  the flux is 
approximately the same for all zenith-angles. 

The predicted LMA enhancement  begins
in the first nighttime bin and the
enhancement is not significantly increased 
when the neutrinos pass through the earth's
core ($\cos(\Theta) > 0.8$) and can even decrease for some parameter
choices. As $\Delta m^2$ decreases, the oscillation length
increases and the effect of averaging over oscillation phases becomes
less effective with the result that some
structure appears in the calculated zenith angle dependence. 
For $\Delta m^2 < 10^{-5} \, {\rm eV^2}$, the departure from a flat
zenith angle dependence becomes relatively large.

Certain solutions of the SMA, those with $\sin^2 2\theta > 0.007$,  predict
a strong enhancement in the core that is not observed and therefore these
SMA solutions are disfavored~\cite{superkamiokande708}.

Figures~\ref{fig:zenith} and \ref{fig:zenithconstm}
show that for 
$\Delta m^2 = (3-4)\times 10^{-5}$ eV$^2$ 
 and a wide range of $\sin^2 2\theta$
the LMA solution provides an excellent
fit to the zenith angle dependence. 

The theoretical uncertainties~\cite{brighter} (due to the
density distribution and composition of the earth and the predicted
fluxes of the standard solar model) are at the level of $0.2$\%, about
an order of magnitude less than the effect that is hinted at by
Fig.~\ref{fig:zenith} and Fig.~\ref{fig:zenithconstm}.

\begin{table}[!t]
\baselineskip=16pt
\centering
\caption[]{Day-Night Asymmetries\protect\label{tab:daynight}}
\begin{tabular}{cccc}
$\Delta m^2$&$A_{\rm n-d}$&$A_{\rm n-d}$&$A_{\rm n-d}$\\
$(10^{-5}~{\rm eV^2})$&$\sin^2 2\theta = 0.9$&$\sin^2 2\theta =
0.8$&$\sin^2 2\theta = 0.7$\\
\noalign{\smallskip}
\hline
\noalign{\smallskip}
8&0.019&0.018&0.017\\
4&0.049&0.050&0.049\\
3&0.069&0.073&0.071\\
2&0.100&0.107&0.107\\
1.6&0.126&0.135&0.136\\
\end{tabular}
\end{table}

\subsection{The Day-Night Spectrum Test}
\label{subsec:spectrumsolar}

The spectral dependence of the day-night effect provides additional and
independent information about neutrino properties that is not
contained in the difference between the total  day and the night event
rates.
The new physical information that is described by the difference
between the day and the night energy spectra is the extent to which
the different energies influence the total rates.

Figure~\ref{fig:daynightspect} compares the calculated 
electron recoil energy spectra for the 
day and the night for a range of values
of $\Delta m^2$ and $\sin^2 2\theta$. Again, we have plotted on the
vertical axis of Fig.~\ref{fig:daynightspect} the
ratio, $R(E)$ [see Eq.~(\ref{eq:Rdefinition})],
of the measured recoil energy spectrum to the spectrum expected if
there is no distortion
\footnote{The average energy spectra during the day and during the night
have been calculated  also in Ref. \cite{marispetcov97}. However,
these authors did not include the SuperKamiokande energy resolution function, 
which leads to a significant smearing effect (see the discussions in
Refs.~\cite{bl96,brighter,guth99}).}. 
The night spectra lie above the day spectra for all relevant
energies. Moreover, the difference of the nighttime and the daytime
rates increases with energy, reflecting the fact that for the LMA
solutions the regeneration effect increases with $\Delta m^2$.

The 
systematic difference between the day and the night spectra 
 is most easily isolated when the difference
due to the average rates is removed from both the day and the night spectra.

Figure~\ref{fig:diffdaynightspect} compares 
the daytime and the nighttime spectra when both spectra are normalized to the
same total rate (the observed SuperKamiokande rate).  
This
equal-normalization removes the difference that is normally referred
to as the `day-night effect'. 
We refer to the equal-normalization 
procedure as  the `day-night spectrum test'.

Figure~\ref{fig:diffdaynightspect}  shows that 
for energies below (above) $10$ MeV, the predicted differential daytime
spectrum is higher (lower) than the predicted 
differential nighttime spectrum.  
The two normalized spectra become equal in rates at about $10$ MeV.
LMA predicts that the nighttime spectrum contains relatively more high
energy electrons than the
daytime spectrum.  

Figure~\ref{fig:diffdaynightspect} shows that, when both spectra are 
normalized to have the same total rates, the average daytime spectrum
between $5$ MeV and 
 $10$ MeV is, for the parameters chosen,
 about  $\sim 1.5 $\% greater than the average nighttime
spectrum. 
The oscillation parameters used to construct
Fig.~\ref{fig:diffdaynightspect} are 
$\Delta m^2 = 2\times 10^{-5}
{\rm eV^2}$ and $\sin^2 2\theta = 0.8$.
We have plotted figures similar to Fig.~\ref{fig:diffdaynightspect}
for a number of LMA solutions.
For a fixed $\Delta m^2$, the equally normalized spectra are all very
similar to each other. 
However, the amplitude of the difference between the
night and the day spectra, about $3$\% at $5$ MeV for the example
shown in Fig.~\ref{fig:diffdaynightspect}, decreases to about $2$\% for
$\Delta m^2 = 4\times 10^{-5}
{\rm eV^2}$ and is only about $0.5$\% for $\Delta m^2 = 8\times 10^{-5}
{\rm eV^2}$ .

Does the day-night spectrum test provide information 
independent of the information
obtained from the integrated day-night rate effect or the more general
zenith-angle effect? Yes, one can imagine that the day-night integrated
effect (cf. Sec.~\ref{subsec:integrated}) 
has been measured to be of order $6$\% 
in agreement with LMA predictions, 
but that when the
day-night spectrum test is applied the appropriately normalized 
 day rate at energies less than $10$ MeV 
lies below the similarly normalized nighttime rate.
This later result would be inconsistent with the
predictions of the LMA solution and therefore would provide
information not available by just measuring the integrated difference
of day and of night rates.

The easiest way to apply the day-night spectrum test  is to
normalize the day and night spectra to the same total number of events
and then 
compare the number of day events below $10$ MeV with the number of
night events below $10$ MeV.
In principle, one could divide the data into a number of 
different bins and test for the similarity to the predicted shape
shown in Fig.~\ref{fig:diffdaynightspect}, but the
Poisson fluctuations would dominate if the data were divided very finely.

The change in slope with energy, 
illustrated in 
Fig.~\ref{fig:diffdaynightspect}, between equally normalized day and night
recoil energy spectra may
 provide a new test of the LMA solution. It will also
be useful to calculate the predicted change in slope 
for other solar neutrino solutions.

\section{Seasonal dependences}
\label{sec:seasonal}

If  the LMA solution is correct, 
seasonal dependences occur as a result
of the same physics that gives rise to the zenith-angle dependence of
the event rates. At nighttime, oscillations in matter can reconvert
$\nu_\mu$ or $\nu_\tau$ to $\nu_e$ or matter interactions can also 
cause the inverse process. The
seasonal dependence arises primarily because the night is longer in the winter
than in the summer. Early discussions of the seasonal effect with
applications to
radiochemical detectors is given
in Refs.~\cite{cribier}. A recent discussion of seasonal effects is
presented in Ref.~\cite{Valle}\footnote{
The principal results given in this section, including  
 Eq.~(\ref{eq:asymmetryidentity}),
have been presented and discussed by A. Yu. Smirnov at the Moriond
meeting for 1999~\cite{smirnovmoriond}.
The seasonal variations found in Ref.~\cite{Valle}
 are in good agreement with our results 
for the same values of the oscillation parameters.}.

\subsection{Predicted versus observed seasonal dependence}
\label{subsec:predictedseasonal}

Figure~\ref{fig:seasonalno} shows the predicted LMA variation of the
total event rates as a function of the season of the year. In
constructing Fig.~\ref{fig:seasonalno},  we corrected  the 
counts for the effect of the
eccentricity of the earth's orbit. We show in
Fig.~\ref{fig:seasonalno} the predictions  for a
threshold of $6.5$ MeV; Fig.~\ref{fig:seasonal11pt5} shows similar
results for a threshold of
$11.5$ MeV. In both cases, the seasonal
effects are small, $\sim 1$\% to $2$\%, although the higher-energy
events exhibit a somewhat  larger ($30$\% or $40$\% larger) variation.
The characteristic errors bars for SuperKamiokande measurements 
three years after beginning the operation 
are larger, typically $\sim
4$\% (cf. Fig.~\ref{fig:seasonal}),
than the predicted
LMA seasonal variations.

How do the SuperKamiokande data compare with the observations? 
For a threshold energy of $6.5$ MeV, the 
data corrected for the earth's eccentricity are not yet
available. Therefore, we compare in Fig.~\ref{fig:seasonal} the
observed~\cite{superkamiokande708} and the predicted seasonal plus
eccentricity dependence. The eccentricity dependence is larger (for the
allowed range of parameters) than
the predicted LMA seasonal dependence; 
hence, the predicted variations in 
Fig.~\ref{fig:seasonal} are larger than in Fig.~\ref{fig:seasonalno}.
The regeneration effect enhances the eccentricity effect in the
northern hemisphere. In the southern hemisphere, the eccentricity and
the  regeneration-seasonal effects have opposite signs. This
difference is in principle detectable; if the seasonal asymmetry is
$7$\% in the northern hemisphere then it would be about $3$\% in the
southern hemisphere.
  
Unfortunately, the existing statistical error bars are too large to show
evidence of either the eccentricity or the LMA seasonal dependence.

If the LMA solution is the correct description of solar neutrino
propagation, it appears likely from Fig.~\ref{fig:seasonal} that 
of order $10$ years of SuperKamiokande 
data will be required in order to see a highly
significant seasonal effect due to LMA neutrino mixing.

\subsection{The relationship between the summer-winter and the
day-night effect}
\label{subsec:swvsdn}

Since the physics underlying the day-night and the seasonal
dependences is the same, \hbox{i.e.},  non-resonant matter mediated
neutrino oscillations, 
there must be a relationship between the two
effects. We outline here a brief derivation of a formula that connects
the size of the day-night asymmetry defined in
Eq.~(\ref{eq:daynightasym}) 
and the winter-summer asymmetry, which we define as

\begin{equation}
A_{\rm W ~-~ S} ~\equiv~ 
2\left({ {\rm Winter - Summer} \over {\rm Winter + Summer} }\right) .
\label{eq:defwintersummer}
\end{equation}
Here `Winter' (`Summer') is the signal averaged over the period from November
15 to February 15 (May 15 to August 15).

Neutrinos reaching the earth from the sun will be in an incoherent
mixture of mass eigenstates~\cite{dighe99,stodolsky98}. 
When the sun is below the horizon, the
presence of electrons in the earth will cause some transitions to
occur between different mass states. By solving the problem of
matter induced neutrino transitions in a constant density medium, one
can see that the characteristic oscillation length in the earth 
is always less than
or equal to the vacuum oscillation length divided by $\sin 2\theta$.
The transition probability is proportional to $\sin^2(\phi)$, where 

\begin{equation}
\phi ~\geq~ {{\pi R \sin 2\theta} \over {L_V} } = 17
 {\sin 2\theta} \left({R \over R_{\rm earth}}\right)
\left({{10 {\rm MeV} } \over {E} }\right)
\left({ {\Delta m^2} \over {2\times 10^{-5}
{\rm eV^2} } } \right),
\label{eq:phidefn}
\end{equation}
where $R$ is the distance traversed in the earth.  For any of the
nighttime bins discussed in Sec.~\ref{sec:zenithdependence}, the phase
of the oscillation  is large so that even relatively small changes in 
distance and  energy will lead to fast oscillatory behavior that
causes the survival probability to average 
to a constant in all of the bins (cf. Eq.~\ref{eq:phidefn}).
This is the reason why the zenith-angle enhancement shown in
Fig.~\ref{fig:zenith} is approximately a constant, independent of
zenith-angle. 

The event during the night and the event rate 
during the day are each approximately 
constant and may be represented  as $R_N$ and $R_D$, respectively. Let
the average duration of night  be $t_S$ hours during the summer and
$t_W$ hours
during the winter. Then the summer signal is proportional to 
$\left[ R_N t_S + R_D(24
- t_S) \right]$.  There is a similar expression for the winter signal:
$\left[ R_N t_W + R_D(24
- t_W) \right]$. Then simple algebra shows that the seasonal asymmetry
is

\begin{equation}
A_{\rm W - S}~=~A_{\rm n -d} 
\left( {{t_W - t_S} \over 24 } \right).
\label{eq:asymmetryidentity}
\end{equation}
Thus the seasonal variations are proportional to the night-day
asymmetry. The larger the day-night effect, the larger is the LMA
predicted seasonal variations due to regeneration.
For the location of SuperKamiokande, $t_W = 13.9$ hr and $t_S = 10.1$
hr, and the parenthetical expression in
Eq.~(\ref{eq:asymmetryidentity}) is about $1/6$.

Equation~(\ref{eq:asymmetryidentity}) gives the approximate relation
between the seasonal and the day-night asymmetries and makes clear why the
seasonal dependence is about $6$ times smaller than the already small
day-night effect. For the central value of the measured range of
day-night asymmetries~\cite{inoue99} (cf. Eq.~\ref{eq:daynightasym}), 
 we find from Eq~(\ref{eq:asymmetryidentity}) $A_{\rm W - S} \sim 1$ \%,
in agreement with our explicit calculations.

\section{Vacuum versus LMA oscillations}
\label{sec:vacuumvsLMAseasonal}

Table~\ref{tab:confrontation} summarizes the most striking 
 differences in the predictions of the LMA and the vacuum oscillation
 solutions of the solar neutrino problems. We now discuss some aspects
 of these differences in more detail.

\begin{table}[!t]
\centering
\caption[]{LMA versus vacuum
oscillations\protect\label{tab:confrontation}.
Section~\ref{sec:zenithdependence} 
contains a quantitative discussion of the day-night
integrated effect, the zenith angle dependence of the rate, and the
day-night spectrum test. Seasonal effects are discussed in
Sec.~\ref{sec:seasonal} and spectrum distortion is discussed in
Sec.~\ref{sec:spectrum}. }
\begin{tabular}{lcc}
\multicolumn{1}{c}{Phenomenon}&LMA&Vacuum\\
\noalign{\smallskip}
\hline
\noalign{\smallskip}
Day-night integrated effect&small but non-zero&zero\\
\noalign{\smallskip}
Zenith-angle dependence of rate&small but non-zero&zero\\
\noalign{\smallskip}
Day-night spectrum test&small but non-zero&zero\\
\noalign{\smallskip}
Spectrum distortion&flat relative spectrum&can explain upturn at 
large energies\\
\noalign{\smallskip}
Seasonal effect&eccentricity dominates&can be large
and energy dependent\\
\end{tabular}
\end{table}

The sharpest distinctions between vacuum and LMA predictions 
are  expressed in the day-night differences. 
The LMA predicts non-zero day-night differences for all three of the
tests discussed in Sec.~\ref{sec:zenithdependence} and listed as
the first three phenomena in Table~\ref{tab:confrontation}. 
Vacuum oscillations
predict zero for all these  day-night phenomena.

The spectrum distortion is predicted to be small for the LMA solution
(cf.~Fig.~\ref{fig:specconsttheta}). 
The ratio of the observed to the
standard spectrum is essentially constant for energies above $6.5$
MeV, although a modest upturn can occur at lower energies. Vacuum
oscillations allow a more varied set of spectral distortions and can
accommodate, without enhanced $hep$ production, the possibly indicated
upturn in the relative spectrum beyond $13$ MeV.

One can use seasonal variations to distinguish between vacuum
oscillations and LMA oscillations. For an early discussion of this
possibility, see Ref.~\cite{krastev95}. The LMA solution predicts that
the seasonal effects are smaller than the geometrical effect arising
from the eccentricity of the earth's orbit. Moreover, there is only a
weak  enhancement
of the LMA seasonal effect with increasing energy threshold.
On the other hand, for vacuum oscillations the seasonal effects can be
comparable with the geometrical effect and there can be a strong
enhancement 
(or an almost complete suppression, even a reversal of the sign) 
of the seasonal effect as the threshold energy is increased
(see first paper in Ref. \cite{krastev95}).

For the LMA solution,
regeneration occurs but only during the night. Therefore, the LMA
solution predicts that there is no
seasonal dependence of the daytime signal beyond 
that expected from the eccentricity of the
earth's orbit.  The vacuum oscillation solutions predict that the day
and the night seasonal dependences will be the same.  Therefore, in
principle one
could  distinguish between vacuum oscillations and LMA oscillations by
comparing the seasonal dependence observed during daytime with the
seasonal dependence observed at night.

The LMA solution predicts for experiments measuring the low energy
$^7$Be neutrino line 
(BOREXINO, KamLAND, LENS)
that there will not be a significant contribution
to the seasonal effect beyond that expected from the eccentricity of
the earth's orbit.  This is because the day-night difference is very
small at low energies in the LMA solution~\cite{brighter,bks98}. 
For vacuum oscillations, there can be significant seasonal dependences
of the $^7$Be line 
in addition to the purely geometrical effect .

SuperKamiokande data  can be used to test for  an
enhancement of the seasonal variations as the energy threshold for
selecting events is increased. Such an enhancement might  be produced
by vacuum oscillations.  Vacuum oscillations 
with $\Delta m^2 \sim 4 \times 10^{-10}~{\rm eV^2}$, which give the
best description of the recoil energy spectrum, 
provide a distinct
pattern for the enhancement: the 
effect of eccentricity is larger than the effect of oscillations
for a threshold energy of $6.5$ MeV, while  the effect of oscillations is
comparable with the eccentricity effect for a threshold of $11.5$
MeV~\cite{krastev95}.

Comparing Fig.~\ref{fig:seasonal11pt5} and
Fig.~\ref{fig:seasonal}, we see that 
the  enhancement with energy threshold predicted by  LMA is
weaker than for vacuum oscillations. For LMA, the increase in the
threshold only enhances the seasonal variations by
 of order $30$\% to $40$\% of an already small effect. This
insensitivity to changes in energy threshold is  
in agreement with
our calculations of the day and night recoil energy spectra
(see Sec.~\ref{subsec:swvsdn}).

Practically speaking, LMA predicts for SuperKamiokande 
that there will be no measurable
change in the seasonal effect with increasing energy threshold.

\section{Summary and discussion}
\label{sec:discussion}

\subsection{The current situation}
\label{subsec:current}

The Large Mixing Angle 
 solution is consistent with all the available data from solar
neutrino experiments. 
Figure~\ref{fig:zoom} shows the allowed region of the LMA 
parameters in the approximation of two oscillating neutrinos.
 
We have investigated in this paper 
the LMA predictions  for SuperKamiokande of the  
distortion of the electron recoil energy spectrum, the 
integrated day-night effect, the 
zenith-angle
dependence of the event rate, and the seasonal dependences. 
We have also evaluated and discussed an independent test, which we call the
day-night spectrum test. 
Finally, we have analyzed the possibilities for distinguishing between
LMA and vacuum oscillations (see Table~\ref{tab:confrontation} and the
discussion in Sec.~\ref{sec:vacuumvsLMAseasonal}).
We previously showed that the LMA solution is globally consistent
with the measured rates in the chlorine, Kamiokande, SAGE, GALLEX, and
SuperKamiokande experiments~\cite{bks98}.

The electron recoil energy spectrum predicted by the LMA solution 
 is practically
undistorted at  energies above $7$ MeV, i.e., the $\nu_e$ 
survival probability is essentially independent of energy at high energies. 
However, the recoil energy spectrum that is
measured by SuperKamiokande suggests an increased rate, 
relative to the standard recoil energy spectrum, for
energies above $13$
 MeV~\cite{superkamiokande300,neutrino98,superkamiokande504,superkamiokande708}. 

We have discussed  in Sec.~\ref{subsec:summaryofdata} 
two  experimental possibilities for explaining
the upturn at large energies of the recoil energy spectrum: (1) a
statistical fluctuation that will go away as the data base is
increased; and (2) an error in the absolute energy
calibration.  
Figure~\ref{fig:energyerror} shows that a best-fit to the upturn at
large energies would require a several sigma error in the absolute
energy calibration.
Both of these seemingly unlikely possibilities, a large statistical
fluctuation and a large error in the energy calibration,  
will be tested by future
measurements with SuperKamiokande.

There are at least two possibilities for explaining the 
spectral upturn that do not involve experimental errors or statistical
fluctuations: (1) a $hep$ flux that is approximately $10$ to $30$
times larger than the conventional nuclear physics estimate; and (2) vacuum
oscillations. We have concentrated in this paper on the explanation
that invokes a larger-than-standard $hep$ flux. Our results are
illustrated in Fig.~\ref{fig:specconsttheta} 
Even when the enhanced rate is used
in the calculations, the $hep$ neutrinos are so rare that they do not
effect anything measurable except the energy spectrum.

At energies below $7$ MeV, the LMA solution predicts a slight upturn
in the recoil energy spectrum relative to the standard model 
energy spectrum. This effect is shown in 
Fig.~\ref{fig:specconsttheta} 
and may be detectable in the future
with much improved statistics.

However, Fig.~\ref{fig:energyerror} shows that even a modest error in
the absolute energy calibration could give rise to an apparently
significant energy distortion. In future analyses, it will be
important to include the absolute energy calibration error as one of
the parameters that is allowed to vary in determining the best fit and
the uncertainty in the spectrum shape.

The zenith-angle dependence and the integrated day-night effect observed by
SuperKamiokande are consistent with the predictions of the LMA and, in
fact, show a hint of an effect ($\sim 1.6\sigma$) that is predicted by LMA
oscillations (see Ref.~\cite{inoue99} and Eq.~\ref{eq:daynightasym}).
Figures~\ref{fig:zenith} and \ref{fig:zenithconstm}
show that the predicted and the observed
nighttime enhancement are both relatively flat, approximately independent of
the zenith angle.  Moreover, all five of the measured nighttime rates
are above the average daytime rate.
However, the results are not very significant
statistically. 
If the LMA description is correct, then 
another $5$ to $10$ years of SuperKamiokande data
taking will be required in
order to reveal a several standard deviation effect.  

The difference
between the daytime and the nighttime recoil energy spectra may be 
detectable in the future. Figure~\ref{fig:daynightspect} shows
separately the predicted
day and the predicted night 
spectral energy distributions. 

The most useful way to test for differences between the shapes of the
daytime and the nighttime energy spectra is to normalize both spectra
to the same value.  Figure~\ref{fig:diffdaynightspect} shows the
predicted difference between the spectra observed during the day and
the spectra observed at night. It will be extremely interesting to
test the hint that a day-night effect has been observed with 
SuperKamiokande by comparing, as in Fig.~\ref{fig:diffdaynightspect},
the equally-normalized day and night spectra. The simplest way of
performing this test would be  to compare the total area of the energy
spectrum that is observed below $10$ MeV at night with the total area
observed below $10$ MeV during the day. The predicted average  
difference 
in the day-night spectrum test is about $1.5$\% for 
$\Delta m^2 = 2\times 10^{-5}
{\rm eV^2}$ and decreases to $\sim 0.5$\% for 
$\Delta m^2 = 8\times 10^{-5}
{\rm eV^2}$.

If the LMA solution is correct, then when  the day
and night spectra are normalized so that they have equal total areas, then the
area under the daytime spectrum curve  below $10$ MeV must be larger  than the
area of the nighttime spectrum  below $10$ MeV. 
No significant difference between the daytime and the nighttime
spectra is expected if vacuum oscillations are the correct
solution of the solar neutrino problems.

Seasonal differences are predicted to be small for the LMA
solution.  Equation~(\ref{eq:asymmetryidentity})
shows that seasonal differences are
expected to be reduced relative to the day-night asymmetry by a factor
of order of $6$ for the location of SuperKamiokande, i.e., 
the difference between the average length of the night in
the winter and the summer divided by $24$ hours. The predicted and the
observed seasonal dependences are shown in Fig.~\ref{fig:seasonalno}
and Fig.~\ref{fig:seasonal}.  If the LMA solution is correct, it will
require many years of SuperKamiokande measurements in order to detect
a statistically significant seasonal dependence.

\subsection{How will it all end?}
\label{subsec:howend}

If there is new physics in the neutrino sector, then experimentalists 
need to provide two different levels of evidence in order to ``solve''
the solar neutrino problems. First, 
measurements must be made 
that are inconsistent with standard electroweak theory at
more than the $3\sigma$ level of significance. Second, diagnostic measurements
must be made that uniquely select a single non-standard neutrino solution. 

Where are we in this program?

We are much of the way toward completing the first requirement in a
global sense. A number of authors have
shown~\cite{bks98,hata94,parke95,robertson} that an
arbitrary linear combination of fluxes from different solar nuclear
reactions, each   with an undistorted neutrino energy spectrum, is
inconsistent at about $3\sigma$ or more 
with a global description of all of the available solar
neutrino data.  The data sets used in these calculations have 
gradually been expanded to include
the results of all five solar neutrino experiments: Homestake,
Kamiokande, SAGE, GALLEX, and SuperKamiokande. The results have become
stronger as more data were added. The precise agreement~\cite{BP98} 
of the calculated sound speeds of
the standard solar model with the measured helioseismological data has
provided a further argument in support of the conclusion that some new
physics is occurring. 

However, we do not yet have a measurement of a smoking-gun
phenomenon, seen in a single experiment,  that is by itself
significant at a many sigma level of significance. This has been the
goal of the current generation of solar neutrino experiments.

What can we say about the possibility of achieving this goal with
SuperKamiokande? There may of course be new physical phenomena that
have not been anticipated theoretically and which will show up with a
high level of significance in the SuperKamiokande experiment. We
cannot say anything about this possibility.

\begin{figure}
\centerline{\epsfxsize=5.5in\epsffile{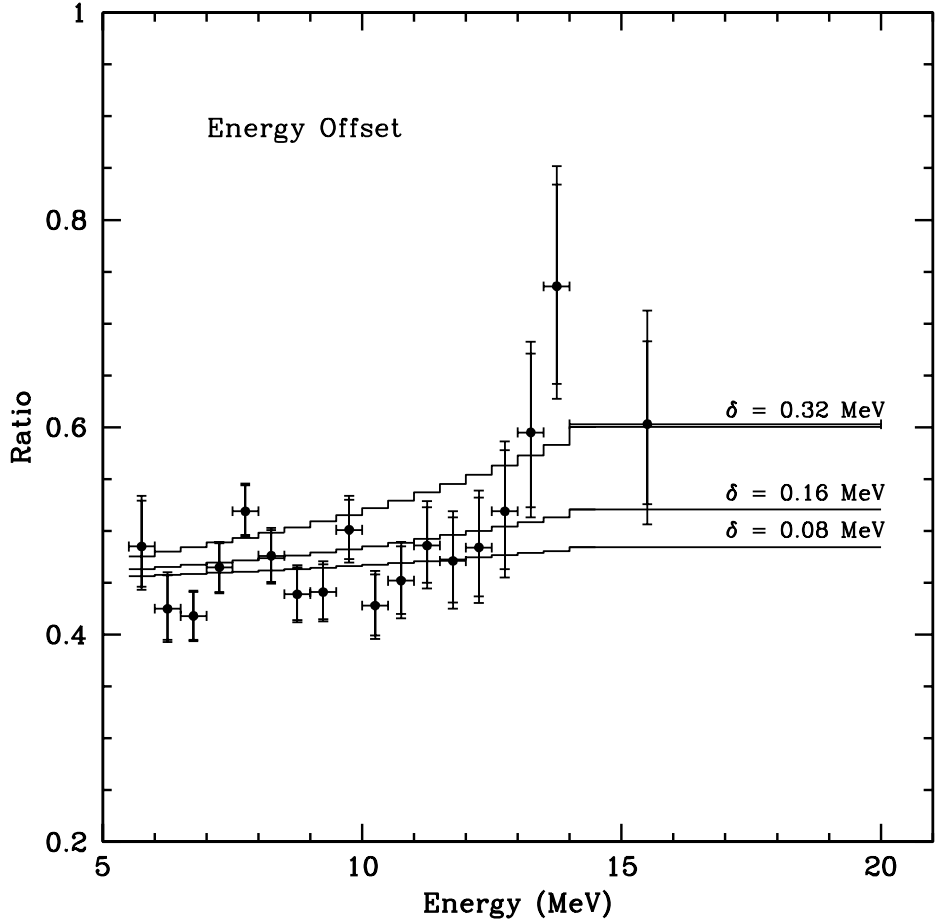}}
\caption[]{ 
The effect of 
a hypothetical   energy calibration error  on the shape of the 
electron recoil energy spectrum.
The data points represent $708$ days of SuperKamiokande 
observations~\cite{superkamiokande708}.
The three curves were calculated from Eq.~(\ref{eq:absolute}) for
constant values of the energy offset parameter, $\delta$, 
of $80$ keV, $160$ keV, and $320$ keV. 
For visual convenience, each curve was calculated with a somewhat
different normalization.
}
\label{fig:energyerror}
\end{figure}

\begin{figure}
\centerline{\epsfxsize=5.5in\epsffile{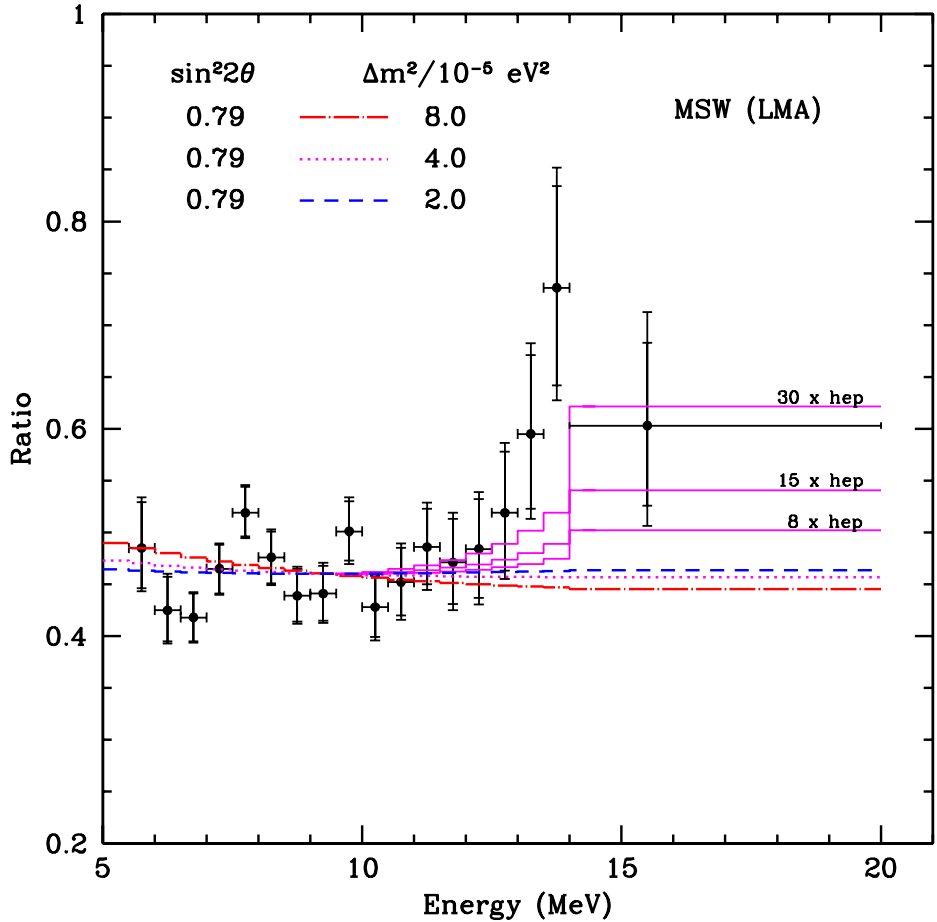}}
\caption[]{Distortion of the $^8$B electron recoil electron spectrum for 
large mixing angle MSW oscillations. The figure
shows the ratio, R, of
the measured number of electrons to the number expected on the basis of
the standard model~\cite{BP98} [see Eq.~(\ref{eq:Rdefinition})].
Solutions are shown with the standard model fluxes of $^8$B and 
$hep$ neutrinos (the approximately horizontal lines) 
and with the standard $hep$ flux multiplied 
(see Ref.~\cite{bk98})
by factors of $8$, $15$,
and $30$ (ratio increases at highest energies). 
The calculated  ratios are shown for three 
different values of $\Delta m^2$ and 
with $\sin^2 2\theta  = 0.79$
 in order to illustrate the range of
behaviors that result from choosing neutrino parameters.
Results similar to Fig.~\ref{fig:specconsttheta} are obtained if 
 $\sin^2 2\theta$ is allowed to vary  over
a representative range of  the allowed global LMA solutions that are
consistent with the average  measured event rates. 
The experimental points show the SuperKamiokande results for 708 days of   
observations, Ref.~\cite{superkamiokande708}. }
\label{fig:specconsttheta}
\end{figure}

\begin{figure}
\centerline{\epsfxsize=5in\epsffile{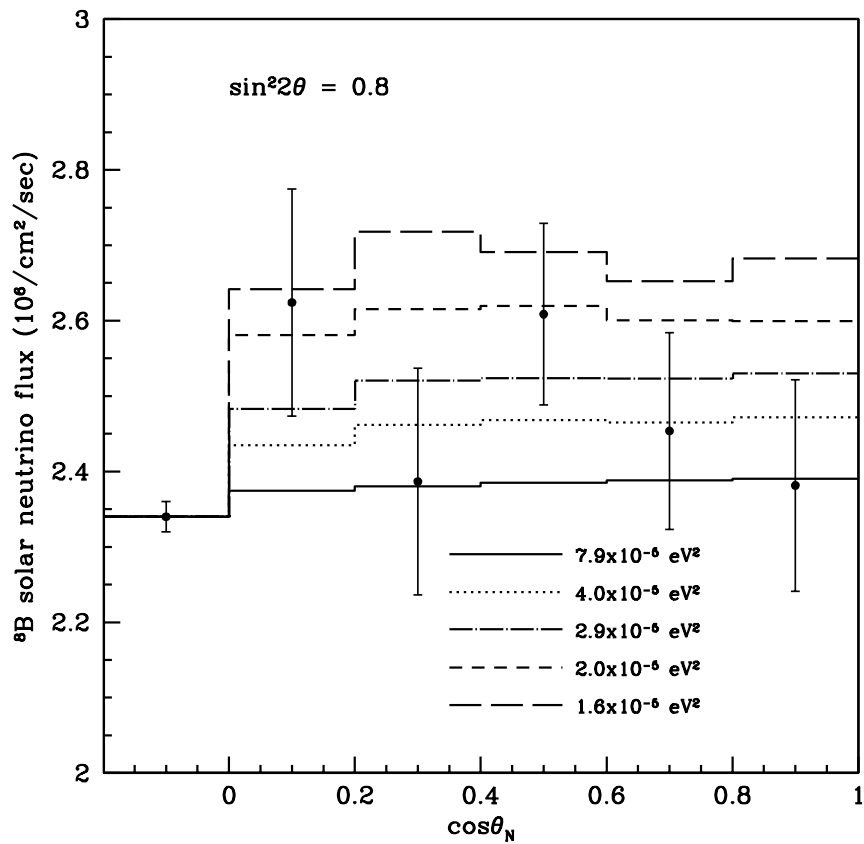}}
\vglue.5truein
\caption[]{ 
The predicted versus the observed zenith angle dependence of the total
event rate above $6.5$ MeV.  
Here $\cos\Theta_N = \cos(\pi - \Theta)$, where $\Theta_N$ is the
nadir angle and $\Theta$ is the zenith angle.  The first bin is the
average of the daytime rate of all of the observations.
The figure shows 
the dependence of the $\nu-e$ scattering
rate in the SuperKamiokande detector upon the nadir  angle of
the sun at the time of the observation is shown for a representative
range of $\Delta m^2$ and fixed $\sin^2 2\theta$.  The results are
averaged over an entire year.
The data are from Ref.~\cite{superkamiokande708}.
  The calculational
procedures are the same as in Ref.~\cite{brighter}. }
\label{fig:zenith}
\end{figure}

\begin{figure}
\centerline{\epsfxsize=5in\epsffile{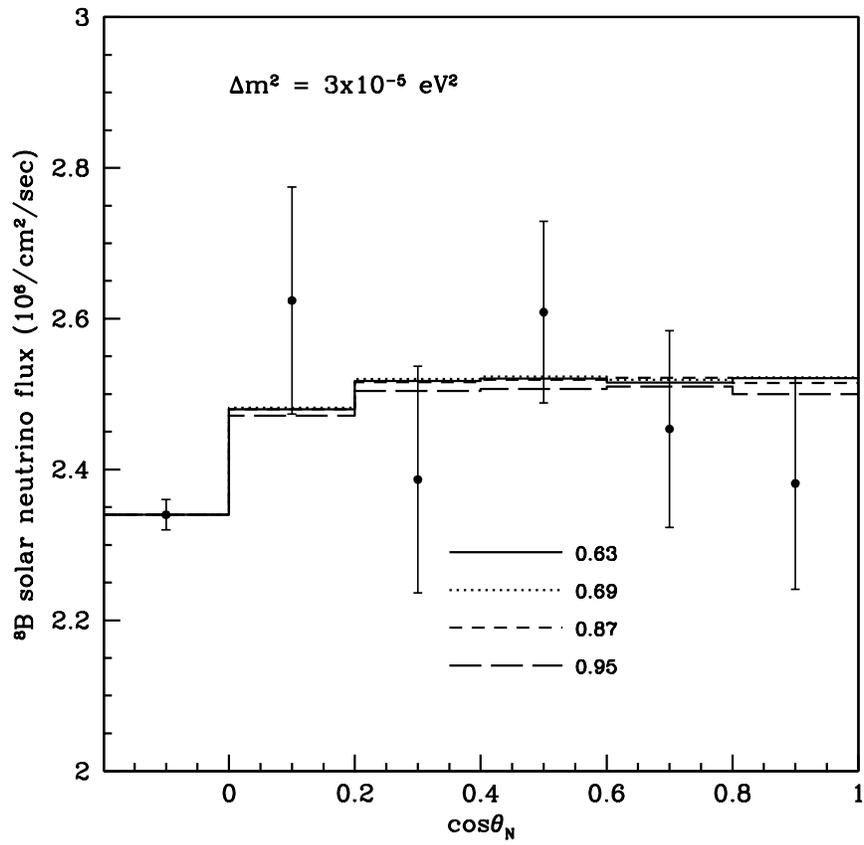}}
\vglue.5truein
\caption[]{ For a fixed value of $\Delta m^2$ and a range of values of
$\sin^2 2\theta$, 
the predicted versus the observed zenith angle dependence of the total
event rate above $6.5$ MeV.  Same as Fig.~\ref{fig:zenith} except for
different oscillation parameters.  This figure makes clear the
difficulty of reducing the error bars so that they are smaller than
the size of the predicted effect.}
\label{fig:zenithconstm}
\end{figure}

\begin{figure}
\centerline{\epsfxsize=5in\epsffile{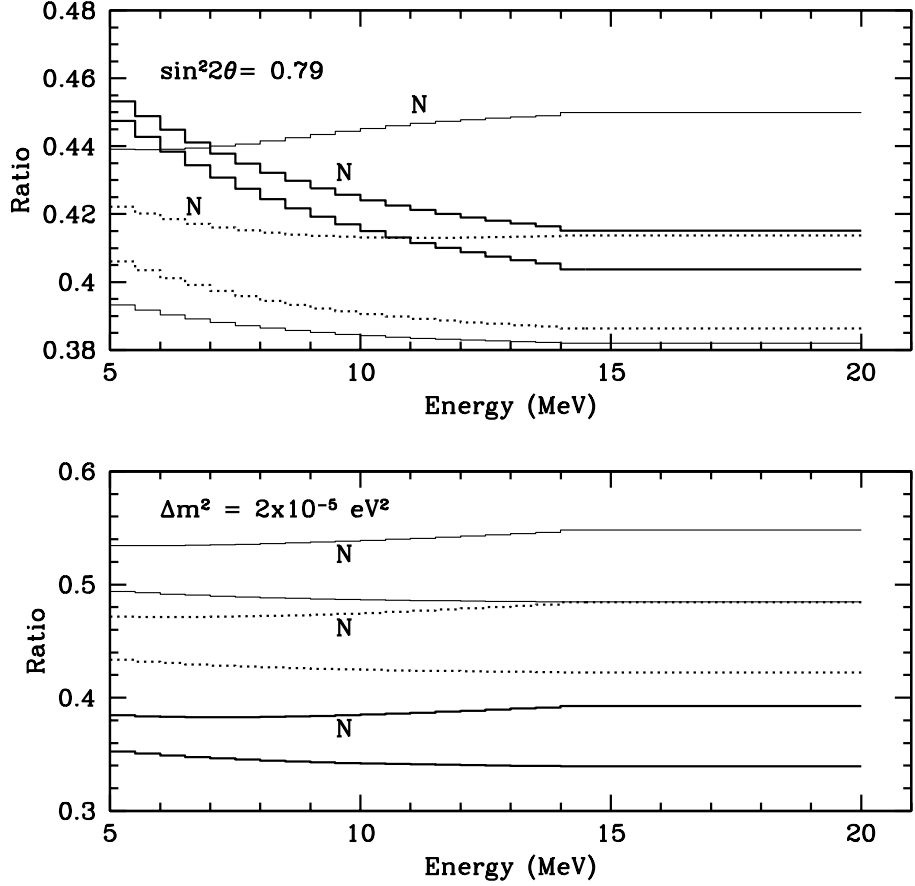}}
\vglue.5truein
\caption[]{ 
The predicted day and night electron recoil energy 
spectra for a representative 
range of $\Delta m^2$ and $\sin^2 2\theta$. The night spectra are
indicated on the figure by $N$; the corresponding day spectra are not
marked. 
For the upper panel, the values of $\Delta m^2$ are (in units of
$10^{-5}{\rm eV^2}$) $7.9$ (thick solid line), $4.0$ (dashed line), and 
$1.6$ (thin solid line). For the lower panel, the values of
$\sin^2 2\theta$ are $0.69$ (thick solid line), $0.87$ (dashed line),
and $0.95$ (thin solid line).}
\label{fig:daynightspect}
\end{figure}

\begin{figure}
\centerline{\epsfxsize=5in\epsffile{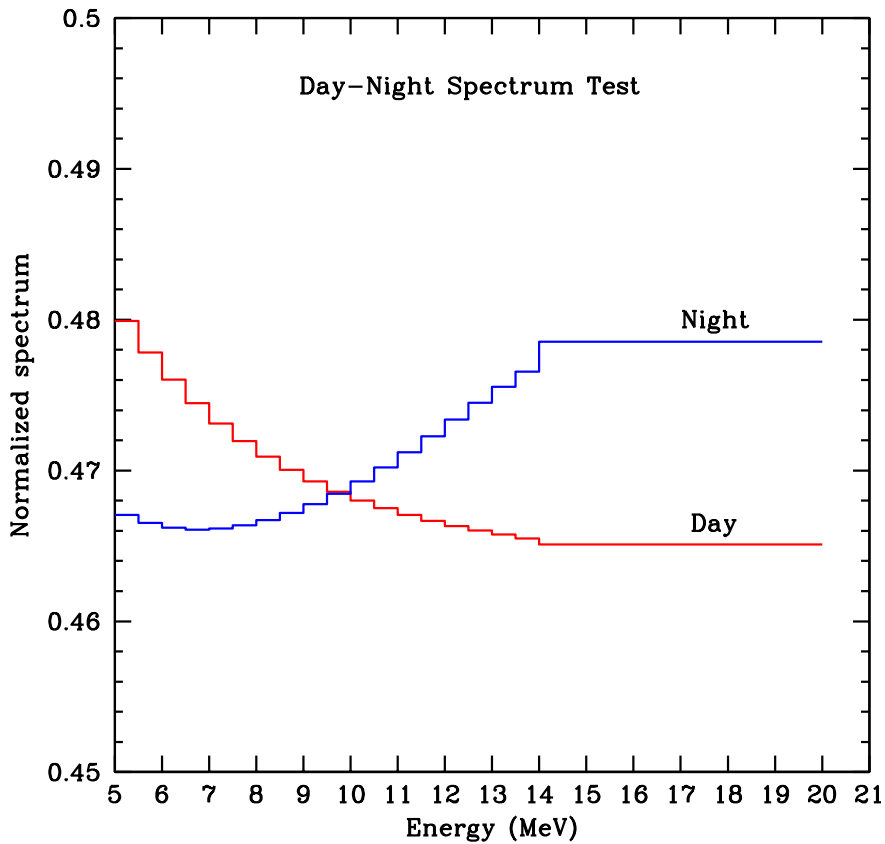}}
\vglue.5truein
\caption[]{ 
The Day-Night Spectrum Test. The figure shows the characteristic
difference between the recoil energy spectrum that is calculated for
daytime observations and the energy spectrum that is calculated for
nighttime observations. The particular curve that is shown in the
figure was calculated for $\Delta m^2 = 2\times 10^{-5}
{\rm eV^2}$ and $\sin^2 2\theta = 0.8$; similarly shaped curves are
obtained for other LMA solutions with the same $\Delta m^2$ but 
different values
of $\sin^2 2\theta$.
The difference between the day and the night spectral shapes decreases
as $\Delta m^2$ increases. }
\label{fig:diffdaynightspect}
\end{figure}

\begin{figure}
\centerline{\epsfxsize=5.5in\epsffile{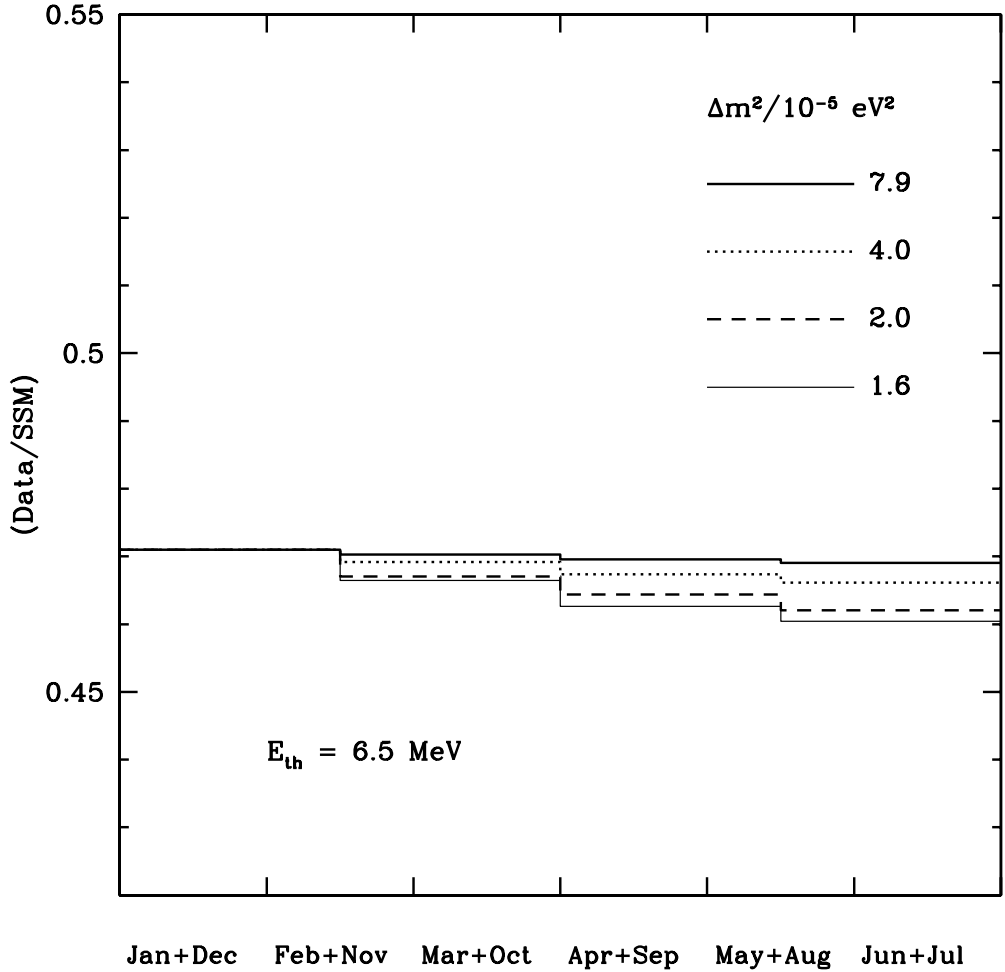}}
\caption[]{ 
The predicted seasonal dependence of the total
event rate with eccentricity removed.  
The dependence of the $\nu-e$ scattering
rate in the SuperKamiokande detector upon the season of the year
is shown for a representative
range of LMA solutions with variable $\Delta m^2$ and
$\sin^2 2\theta = 0.8$.
In order to isolate the seasonal effect of neutrino oscillations, the
effect of the eccentricity of the earth's orbit has been removed. 
Figure~\ref{fig:seasonalno} refers to all events above $6.5$ MeV.}
\label{fig:seasonalno}
\end{figure}

\begin{figure}
\centerline{\epsfxsize=5.5in\epsffile{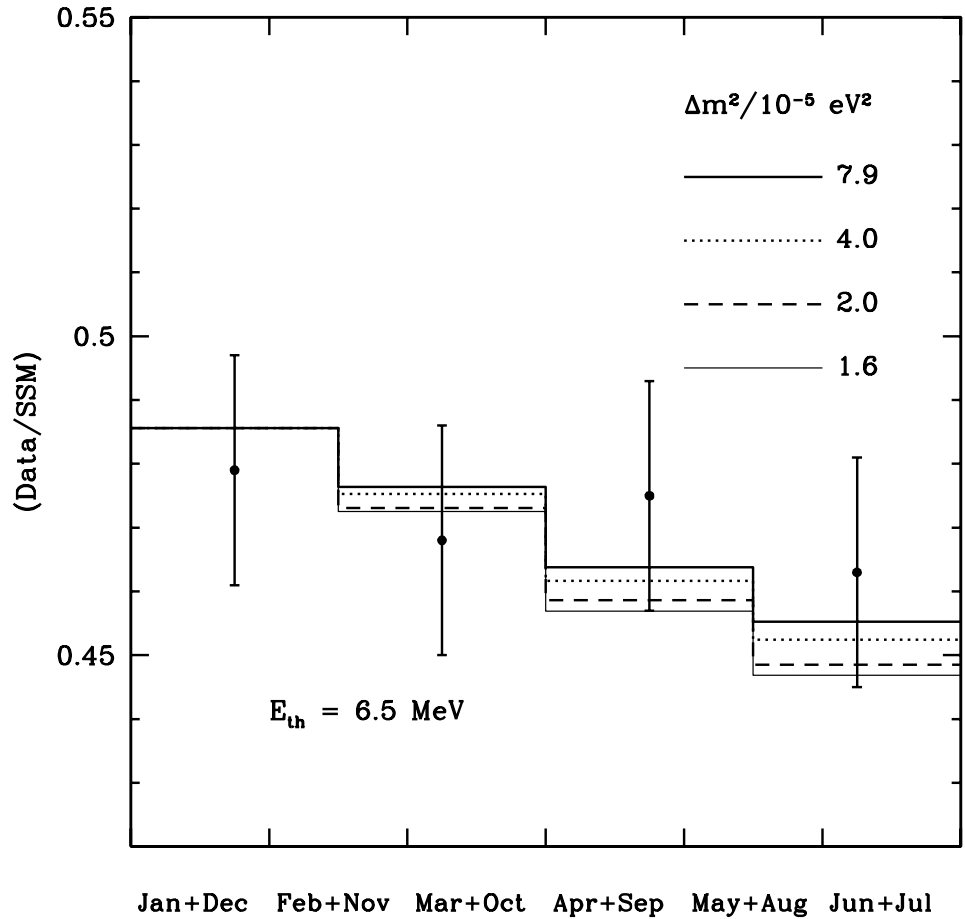}}
\caption[]{ 
The predicted seasonal plus eccentricity dependences of the total
event rate.  
This figure is the same as Fig.~\ref{fig:seasonalno} except that 
for the present figure we have not removed the 
effect of the eccentricity of the earth's orbit
in the predictions or the observations. 
The data are from Ref.~\cite{superkamiokande708}. }
\label{fig:seasonal}
\end{figure}

\begin{figure}
\centerline{\epsfxsize=5.5in\epsffile{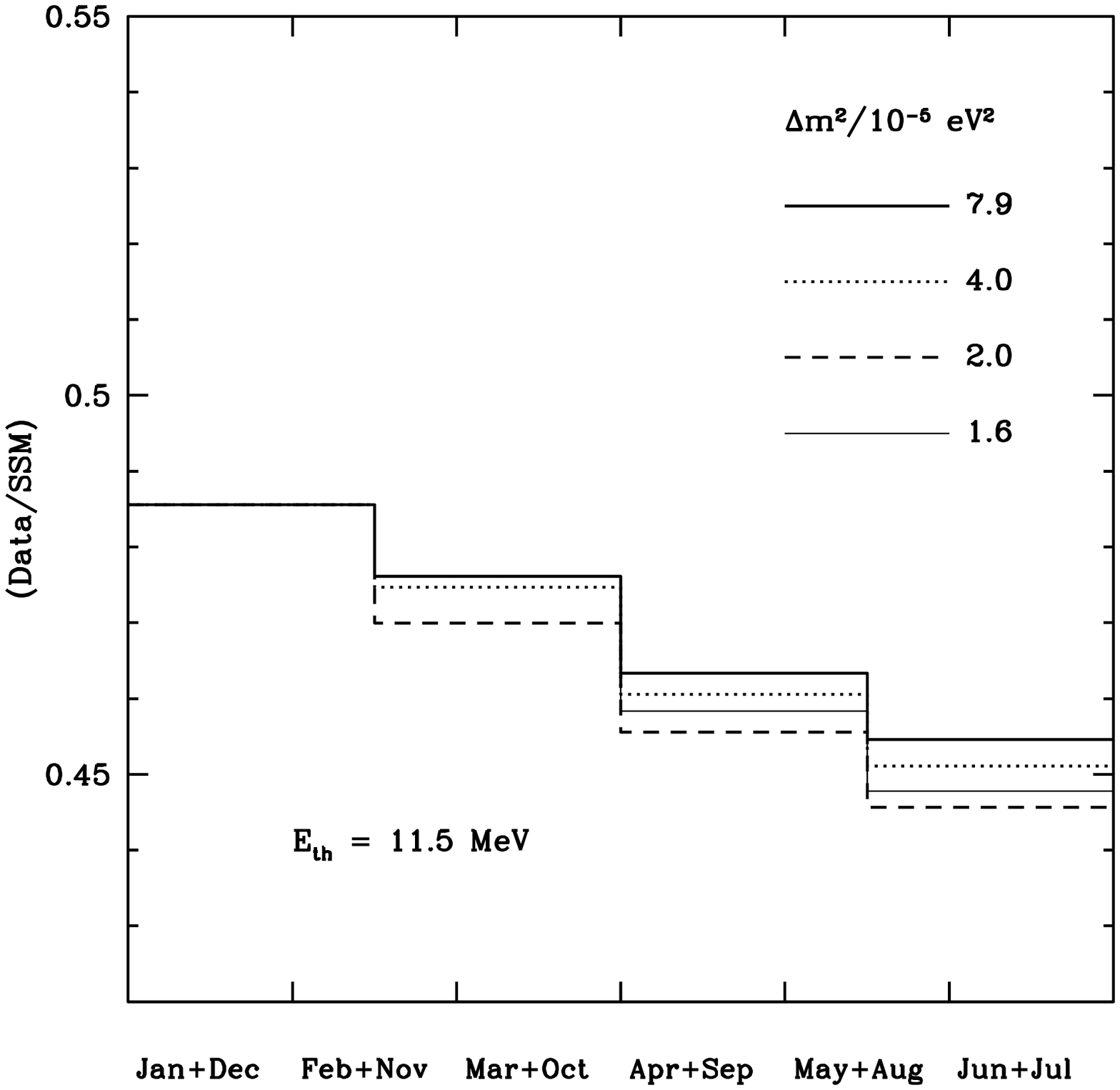}}
\caption[]{ 
The predicted seasonal plus eccentricity dependences of the total
event rate for $E_{\rm th} = 11.5$ MeV.  
This figure is the same as Fig.~\ref{fig:seasonal} except that the
threshold recoil electron energy is larger.}
\label{fig:seasonal11pt5}
\end{figure}

If the LMA description of solar neutrinos is correct, we can use the
results shown in Fig.~\ref{fig:energyerror}--Fig.~\ref{fig:seasonal11pt5} 
to estimate how long
SuperKamiokande must be operated in order to reduce the errors so that
a single effect (spectrum distortion, zenith-angle dependence, or
seasonal dependence) is significant at more than the $3\sigma$ level.
Figures~\ref{fig:energyerror}--\ref{fig:seasonal11pt5} 
show that the errors must be
reduced for any one phenomenon by at least a factor of two in order to
reach a multi-sigma level of significance. Since the available data
comprise $708$ days of operation, over a calendar period of order
three years, it seems likely that
SuperKamiokande will require of order a decade of running in order 
to isolate a single direct proof of
solar neutrino oscillations, provided the LMA description is correct.

Fortunately, 
SuperKamiokande can make a global test of the standard electroweak
hypothesis that nothing happens to neutrinos after they are created in
the center of the sun. This global test 
could  become significant at the several sigma level
within  a few years even if LMA is correct. For example, 
the combined effect of the
zenith-angle dependence plus a possible spectral distortion at low
energies might show up as a clear signal.  Or, both the integrated 
day-night effect
and the day-night spectrum test could be observed.
The combined statistical significance of several difference tests
could be very powerful.
With the great precision and the high statistical significance of the
SuperKamiokande data, we think that there is a good chance that
a many sigma result may be obtainable in a relatively few years. 

{\it Note added in proof.} Recently, the 
SuperKamiokande collaboration has made available  the data 
for $825$ days of observations (T. Kajita, talk given at 
the Second Int.~Conf. ``Beyond the Desert,'' Castle Ringberg, Tegernsee, 
Germany, June 6 - 12, 1999). In this more complete data set, the 
 significance of
the excess of events at  high energies has further decreased and the
Day-Night asymmetry has increased to approximately the $2\sigma$
level. Both of these developments strengthen slightly the case for the LMA
solution. 

\acknowledgments
We are grateful to the  SuperKamiokande
collaboration for continuing to make 
 available their superb preliminary data and for many stimulating discussions.
We are also grateful to David Saxe for skillfully performing some numerical
calculations.
JNB  acknowledges support from NSF grant No. PHY95-13835 
and
PIK acknowledges support from  NSF grant No. PHY95-13835  and
NSF grant No. PHY-9605140.

\newpage

\begin{thebibliography}{99}
\bibitem{chlorine}B. T. Cleveland {\it et al.}, Astrophys. J. {\bf
496}, 505 (1998); B. T. Cleveland {\it et al.}, 
Nucl. Phys. B (Proc. Suppl.) {\bf 38}, 47 (1995); R. Davis,
Prog. Part. Nucl. Phys. {\bf 32}, 13 (1994).
\bibitem{kamiokande}KAMIOKANDE Collaboration, Y. Fukuda {\it et al.}, 
Phys. Rev. Lett. {\bf 77}, 1683 (1996).
\bibitem{SAGE}SAGE Collaboration, V. Gavrin {\it et al.}, in {\it
Neutrino 96}, Proceedings of 
the XVII International
Conference on Neutrino Physics and Astrophysics, Helsinki, edited by  
K. Huitu, K. Enqvist and J. Maalampi 
(World Scientific, Singapore, 1997), p. 14;
V. Gavrin {\it et al.}, in {\it Neutrino 98}, Proceedings of the XVIII International Conference on
Neutrino Physics and Astrophysics, Takayama, Japan, 4--9 June 1998,
edited by Y. Suzuki and Y. Totsuka. To be published in Nucl. Phys. B
(Proc. Suppl.).
\bibitem{GALLEX} GALLEX Collaboration, P.~Anselmann {\it et al.}, 
Phys.~Lett. B {\bf 342}, 440 (1995); GALLEX Collaboration,
W. Hampel {\it et al.},  Phys.~Lett. B 
{\bf 388}, 364 (1996).
\bibitem{superkamiokande300} SuperKamiokande Collaboration, Y. Fukuda {\it
et al.} Phys. Rev. Lett. {\bf 81}, 1158 (1998); Erratum-{\it ibid.} 
{\bf 81}, 4279 (1998).
\bibitem{neutrino98} SuperKamiokande Collaboration, Y. Suzuki, in 
{\it Neutrino 98}, Proceedings of the XVIII International Conference on
Neutrino Physics and Astrophysics, Takayama, Japan, 4--9 June 1998,
edited by Y. Suzuki and Y. Totsuka. To be published in Nucl. Phys. B
(Proc. Suppl.).
\bibitem{superkamiokandeDN} SuperKamiokande Collaboration, Y. Fukuda {\it
et al.} Phys. Rev. Lett. {\bf 82}, 1810 (1999).
\bibitem{superkamiokande504} SuperKamiokande Collaboration, Y. Fukuda {\it
et al.}, Phys. Rev. Lett. {\bf 82}, 2430 (1999).
\bibitem{superkamiokande708} SuperKamiokande Collaboration, M. Smy,
hep-ex/9903034.
\bibitem{suzuki99}Y. Suzuki, talk given at the 17th Intl. Workshop on Weak
Interactions and Neutrinos, Win-99, Cape Town, South Africa,
(unpublished) Jan. 24-30 (1999).
\bibitem{minikata} H. Minakata and  H. Nunokawa,
Phys. Rev. D {\bf 59}, 073004 (1999).
\bibitem{bks98}J. N. Bahcall, P. I. Krastev, A. Yu. Smirnov,
Phys. Rev. D {\bf 58}, 096016 (1998).
\bibitem{BP98}J. N. Bahcall, S. Basu, and M. H. Pinsonneault,
  Phys. Lett. B {\bf 433}, 1 (1998)
\bibitem{msw}L. Wolfenstein, Phys. Rev. D {\bf 17}, 2369 (1978);
S. P. Mikheyev and A. Yu. Smirnov, Yad. Fiz. {\bf 42}, 1441 (1985)
[Sov. J. Nucl. Phys. {\bf 42}, 913 (1985)]; Nuovo 
Cimento C {\bf 9}, 17 (1986).
\bibitem{inoue99}K. Inoue, VIII Intl. Workshop on ``Neutrino Telescopes" 
Venice, February 23 - 26, 1999 (unpublished).
\bibitem{fukuda98}Y. Fukuda et al., Phys. Rev. Lett. {\bf 81}, 1562 (1998).
\bibitem{smirnov99}O.L.G. Peres and A. Yu. Smirnov,  hep-ph/9902312, to
be published in Phys. Lett. B (1999).
\bibitem{balisi}  J.~N.~Bahcall, E.~Lisi, D.~E.~Alburger, L.~De Braeckeleer,
S.~J.~Freedman, and J.~Napolitano, Phys.\ Rev.\ C {\bf  54}, 411 (1996).
\bibitem{bahcall91}J. N.~Bahcall, Phys.~Rev. D {\bf 44}, 1644
(1991).
\bibitem{sirlin}J.~N.~Bahcall, M.~Kamionkowski, and A.~Sirlin, 
Phys.\ Rev.\ D {\bf 51}, 6146 (1995). 
\bibitem{bk98}J. N. Bahcall and P. Krastev, Phys. Lett. B {\bf 436},
243 (1998)
\bibitem{escribano}R. Escribano, J.M. Frere, A. Gevaert, D. Monderen
Phys.Lett. B {\bf444}, 397 (1998).
\bibitem{fiorentini}G. Fiorentini et al. Phys. Lett. B {\bf 444} 387 (1998).
\bibitem{bl96}J. N. Bahcall and E. Lisi, Phys. Rev. D {\bf 54}, 5417
(1996); J. N. Bahcall, P. I. Krastev, and E. Lisi, Phys. Rev. D {\bf
55}, 494 (1997).
\bibitem{liu} Q. Y. Liu, talk given at 17 International Workshop on Weak
Interactions
and Neutrinos, Cape Town, South Africa, January 24--30, 1999.
\bibitem{adelberger98}E. Adelberger et al., Rev. Mod. Phys. {70}, 1265
(1998).
\bibitem{earthreg} S. P. Mikheyev and A. Yu. Smirnov, in {\it '86
Massive Neutrinos in Astrophysics and in Particle Physics},
proceedings of the Sixth Moriond Workshop, edited by O. Fackler and
Y. Tr\^an Thanh V\^an (Editions Fronti\`eres, Gif-sur-Yvette, 1986),
p. 355;
J. Bouchez {\it et al.}, Z. Phys. C {\bf 32}, 499 (1986); M. Cribier,
W. Hampel, J. Rich, and D. Vignaud, Phys. Lett. B {\bf 182}, 89 (1986);
M. L. Cherry and K. Lande, Phys. Rev. D {\bf 36}, 3571 (1987);
S. Hiroi, H. Sakuma, T. Yanagida, and M. Yoshimura, Phys. Lett. B {\bf
198}, 403 (1987); S. Hiroi, H. Sakuma, T. Yanagida, and M. Yoshimura,
Prog. Theor. Phys. {\bf 78}, 1428 (1987); A. Dar, A. Mann, Y. Melina,  
and D. Zajfman, Phys. Rev. D {\bf 35}, 3607
(1988); M. Spiro
and D. Vignaud, Phys. Lett. B {\bf 242}, 279 (1990);
A. J. Baltz and J. Weneser, Phys. Rev. D {\bf 35}, 528 (1987);
A. J. Baltz and J. Weneser, Phys. Rev. D {\bf 37}, 3364 
(1988).
\bibitem{brighter}J. N. Bahcall and P. I. Krastev, Phys. Rev. C {\bf 56},
2839 (1997).
\bibitem{marispetcov97} M. Maris and S. T. Petcov, 
Phys. Rev. D {\bf 56}, 7444 (1997).
\bibitem{guth99}A. Guth, L. Randall, and M. Serna, hep-ph/9903464. 
\bibitem{cribier}M. Cribier et al, Phys. Lett. B {\bf 188} 169 (1987);
M. L. Cherry and K. Lande, Phys. Rev. D {\bf 36} 3571 (1987);
see also A. J. Baltz and J. Weneser, Phys.Rev D {\bf 35} 528 (1987).
\bibitem{Valle} P. C. de Holanda, C. Pena-Garay, M. C. 
Gonzalez-Garcia and J. W. F. Valle, hep-ph/9903473 .
\bibitem{smirnovmoriond}A. Yu. Smirnov, XXXIVth Recontres de Moriond,
Electroweak Interactions and Unified Theories, Les Arcs, March 13--20,
1999; S. P. Mikheyev and A. Yu. Smirnov, in 
{\it New and Exotic Phenomena}, Proceedings of 7th Moriond Workshop, 
ed by O. Fackler and J. Tran Thanh Van, (Editions Frontieres, 1987) p.403;   
and  in {\it Neutrinos} Ed. H. V. Klapdor, (Springer-Verlag, 1987)
p.239,
see viewgraphs available at http://www.lal.in2p3.fr/CONF/Moriond .
\bibitem{dighe99}A. Dighe, Q. Y. Liu, and A. Yu. Smirnov,
hep-ph/9903329 (1999).
\bibitem{stodolsky98} L. Stodolsky, Phys. Rev. D {\bf 58}, 036006 (1998).
\bibitem{krastev95}P.I.Krastev and S.T.Petcov, Nucl. Phys. B {\bf
449}, 605 (1995); S. P. Mikheyev and A. Yu. Smirnov Phys. 
Lett. B429 (1998) 343; 
S. L. Glashow, P. J. Kernan and L. Krauss, hep-ph/9808470;
V. Barger and K. Whisnant, hep-ph/9812273, hep-ph/9903262;  
M. Maris, S.T. Petcov, hep-ph/9903303;
V. Berezinsky, G. Fiorentini,  M. Lissia,  hep-ph/9904225. 
\bibitem{hata94}N. Hata, S. Bludman, and P. Langacker, Phys Rev D {\bf
49}, 3622 (1994).
\bibitem{parke95}S. Parke, Phys. Rev. Lett. {\bf 74}, 839 (1995).
\bibitem{robertson}K. M. Heeger and R. G. H. Robertson, Phys Rev Lett
{\bf 77}, 3720 (1996).
\end{thebibliography}
\end{document}